\documentclass[prb,twocolumn,aps,showpacs,fixfloats]{revtex4}
\usepackage{graphicx}
\usepackage{bm}
\usepackage{amsmath,amssymb}
\usepackage{subfigure}
\usepackage{float}
\usepackage{latexsym}
\usepackage{epstopdf}
\usepackage{color}
\usepackage{enumerate}

\begin{document}
\newcommand{\s}{\scriptscriptstyle}
\newcommand{\uu}{\uparrow \uparrow}
\newcommand{\ud}{\uparrow \downarrow}
\newcommand{\du}{\downarrow \uparrow}
\newcommand{\dd}{\downarrow \downarrow}
\newcommand{\ket}[1] { \left|{#1}\right> }
\newcommand{\bra}[1] { \left<{#1}\right| }
\newcommand{\bracket}[2] {\left< \left. {#1} \right| {#2} \right>}
\newcommand{\vc}[1] {\ensuremath {\bm {#1}}}
\newcommand{\tr}{\text{Tr}}

\title{ Slow dynamics of spin pairs in random hyperfine field:
Role of inequivalence  of electrons and holes
in organic magnetoresistance}

\author{R. C. Roundy  and M. E. Raikh}
\affiliation{Department of Physics and Astronomy, University of Utah, Salt Lake City, UT 84112}

\begin{abstract}
In an external magnetic field B, the spins of the electron and hole
will precess in effective fields ${\bm b}_{\s e} + {\bm B}$ and
${\bm b}_{\s h} + {\bm B}$, where ${\bm b}_{\s e}$ and ${\bm b}_{\s h}$
are random hyperfine fields acting on the electron and hole, respectively.
For sparse ``soft" pairs the magnitudes of these effective fields coincide.
The dynamics of precession for these pairs acquires a slow component, which
leads to a slowing down of recombination.  We study the effect of soft
pairs on organic magnetoresistance, where slow recombination translates
into blocking of the passage of current.  It appears that when ${\bm b}_{\s e}$
and ${\bm b}_{\s h}$ have  identical gaussian distributions the contribution
of soft pairs to the current does not depend on $B$.  Amazingly, small
inequivalence in the rms values of $b_{\s e}$ and $b_{\s h}$ gives rise to a
magnetic field response, and it becomes progressively stronger as the inequivalence
increases. We find the expression for this response
by performing the averaging over ${\bm b}_{\s e}$, ${\bm b}_{\s h}$ analytically.
Another source of magnetic field response in the regime when current is  dominated by soft pairs is
inequivalence of the $g$-factors of the pair partners. Our analytical calculation
indicates that for this mechanism the response has an opposite sign.
\end{abstract}

\maketitle

\section{Introduction}

Due to complex structure of organic semiconductors and their spatial
inhomogeneity it is nearly impossible to identify a unique scenario of
current passage through them. In view of this, it is remarkable that
sizable change of current through a device based on organic
semiconductor takes place in  weak external magnetic fields.
This effect, called organic magnetoresistance (OMAR), seems to be robust,
i.e. weakly sensitive to the device parameters.  Although the first reports
on the observation of organic magnetoresistance (OMAR) appeared decades
ago\cite{frankevich,frankevich1}, systematic experimental study of this
effect started relatively recently. \cite{Markus0,Markus1,Markus2,Markus3,Gillin,Valy0,Valy1,Valy2,Valy3,Blum,Bobbert1,Bobbert2,Wagemans2,Wagemans,Wagemans1}
(see also the review Ref. \onlinecite{WagReview}).

On the theory side, it is now commonly accepted that
the origin of OMAR lies in random hyperfine fields created
by nuclei
surrounding the carriers (polarons).
More specifically, the basic unit responsible for
OMAR is a pair of sites hosting carriers (polarons); the spin state of the pair
is described by the Hamiltonian
\begin{equation}
\label{Hamiltonian}
\widehat{H}={\bm \Omega}_1\cdot \widehat{{\bm S}}_{\s e}+{\bm \Omega}_2\cdot \widehat{{\bm S}}_{\s h}.
\end{equation}
Here $\widehat{{\bm S}}_{\s e}$ are $\widehat{{\bm S}}_{\s h}$ are the spin operators
of the pair-partners (we will assume that they are electron and hole, respectively);
${\bm \Omega}_1={\bm B}+{\bm b}_{\s e}$ and ${\bm \Omega}_2={\bm B}+{\bm b}_{\s h}$
are the {\em full} fields acting on
the spins. They represent the sums of external, ${\bm B}$, and respective
hyperfine fields, ${\bm b}_{\s e}$  and ${\bm b}_{\s h}$.
As was first pointed out by Schulten and Wolynes \cite{Schulten},
due to the large number of nuclei surrounding each pair-partner
and  their slow dynamics,  ${\bm b}_{\s e}$ and ${\bm b}_{\s h}$
can be viewed as classical random fields with gaussian distributions.

In order to give rise to OMAR the Hamiltonian Eq. (\ref{Hamiltonian}) is not
sufficient. It should be complemented by  some mechanism through which
the pair-partners ``know" about each other, so their motion is
correlated  without direct interaction.
The simplest example of such a mechanism
is spin-dependent recombination, i.e. the requirement that electron
and hole can recombine only if their spins are in the singlet, $S$,
state. Then the essence of OMAR can be crudely understood  as a
redistribution of portions of singlets and triplets upon
increasing $B$. This redistribution affects the net
recombination rate. Clearly, the characteristic $B$ for this
redistribution is $\sim b_{\s e}, b_{\s h}$.

\begin{figure}[t]
\includegraphics[width=77mm, clip]{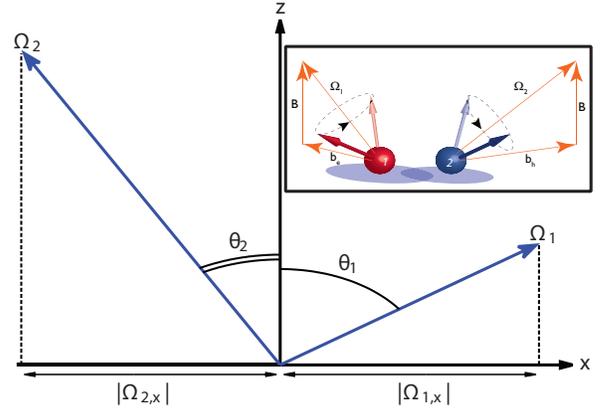}
\caption{Preferential coordinate system used for analysis of the dynamics of
the spin pair. Both fields ${\bm \Omega}_1$, ${\bm \Omega}_2$ reside in the $xz$-plane.
The direction of the quantization axis, $z$, is fixed by the condition
${\bm \Omega}_{1,x}=-{\bm \Omega}_{2,x}$.}
\label{figCoords}
\end{figure}

Naturally, the specific relation between the current and recombination
rate involves also the rate at which the pairs are created.
It is important, though,  that the latter process is not spin-selective.

Existing theories of OMAR can be divided into two groups
which we will call ``steady-state" and ``dynamical".
The theories of the first group\cite{Prigodin} appeared earlier.
In a nutshell (see Ref. \onlinecite{Wagemans} for details), in these theories
the right-hand-side of the equation of motion for the density matrix
$i\dot{\rho}=[\widehat{H}, \rho]$ with Hamiltonian Eq. (\ref{Hamiltonian})
is complemented with ``source" and {\em spin-selective} ``sink" terms.
After that, $\dot{\rho}$ is set to zero. In Refs. \onlinecite{Bobbert1}
current is expressed via the steady-state $\rho$ and subsequently averaged
numerically over realizations of hyperfine fields.

The ``steady-state" approach applies when the pair does not perform many beatings
between $S$ and $T$ during its lifetime, since the beating {\em dynamics} is
excluded by setting $\dot{\rho}=0$.

This beating dynamics has been incorporated into the OMAR theory Ref. \onlinecite{Flatte1},
which appeared last year. This theory  relies on decades old findings in
the field of dynamic spin-chemistry\cite{Schulten,reviews}. Below we briefly
summarize these findings.

If an isolated pair is initially in $S$, it was shown in Ref. \onlinecite{Schulten}
that the averaged probability to find it in $T$ after time $t$ is given either by
the function
\begin{equation}
\label{schultenbig}
p_{\s ST}(t) = \frac{1}{2} \left(1 - e^{-b_{\s e}^2 t^2} e^{-b_{\s h}^2 t^2} \right),
\end{equation}
for strong fields $B \gg b_{\s e}, b_{\s h}$, or by
\begin{multline}
\label{schultensmall}
p_{\s ST}(t) = \frac{3}{4} \left\{
1 - \left[\frac{1}{3}\left( 1 + 2e^{-b_{\s e}^2 t^2}
  -4 b_{\s e}^2 t^2 e^{-b_{\s e}^2 t^2}\right) \right] \right. \\
  \times \left. \left[\frac{1}{3}\left( 1 + 2e^{-b_{\s h}^2 t^2}
  -4 b_{\s h}^2 t^2e^{-b_{\s h}^2 t^2}\right) \right]
\right\},
\end{multline}
for $B \ll b_{\s e}, b_{\s h}$. Here $b_{\s e}$, $b_{\s h}$
are the rms hyperfine fields for electron and hole.
Naturally, the probability approaches $3/4$ at small $B$ and $1/2$ at large $B$.

In the theory of Ref. \onlinecite{Flatte1} the $B$-dependent dynamics
described by Eqs. (\ref{schultenbig}), (\ref{schultensmall})
translates into the $B$-dependent resistance (OMAR) on the basis
of the following reasoning. The dynamics $p_{\s ST}(t)$
leads to prolongation of the recombination time
(hopping time, $\tau_{\s h}$, in the language of Ref. \onlinecite{Flatte1}). This
prolongation is quantified by

\begin{equation}
\label{arrow}
\frac{1}{\tau_{\s h}} \rightarrow
\frac{1}{\tau_{\s h}} \int dt (1 - 3 p_{\s ST}(t))e^{-t/\tau_{\s h}}.
\end{equation}
The meaning of Eq. (\ref{arrow}) is that a pair should
stay in $S$ in order for a hop to take place.
Prolongation of hopping time leads to a $B$-dependent increase
of the resistance.
The authors of Ref. \onlinecite{Flatte1} evaluated  $p_{\s ST}(t)$
for arbitrary $B$, while in calculation of OMAR they
assumed that bare hopping times, $\tau_{\s h}$, have an exponentially
broad distribution.

Both theories Refs. \onlinecite{Wagemans}, \onlinecite{Flatte1}
take as a starting point a pair with the Hamiltonian Eq. (\ref{Hamiltonian})
describing its spin states and preferential recombination (hopping) from $S$.
The dynamics of this seemingly simple entity, which is crucial for OMAR,
possesses some nontrivial regimes.
Uncovering these regimes is a central goal for the present paper.
The other goal is to demonstrate that nontrivial dynamics can
manifest itself in OMAR.

To underline that the spin dynamics of two carriers in
non-collinear magnetic fields which can recombine
only from $S$ can be highly nontrivial, we note that
separation of this dynamics into $S$-$T$ ``beating" stage
followed by instantaneous hopping after time $\tau_{\s h}$,
as in theory \onlinecite{Flatte1}, is not always possible.
It is quite nontrivial that spin-selective recombination of carriers
can exert a {\em feedback} on the spin dynamics. As an
illustration of  this delicate issue we invoke the example of
cooperative photon emission discovered by R. H. Dicke\cite{Dicke}.
In the Dicke effect one superradiant state of a group of emitters
having a very short lifetime automatically implies that all the
remaining states are subradiant and have {\em anomalously long}
radiation times. Below we demonstrate that a similar situation
is realized in dynamics of two spins when recombination from
$S$ is very fast. We will see that the remaining $3$  modes
of the collective spin motion become very ``slow".

Our analysis reveals the exceptional role of the ``soft"
pairs, which are sparse configurations of ${\bm b}_{\s e}$, ${\bm b}_{\s h}$
for which full fields ${\bm \Omega}_1$, ${\bm \Omega}_2$ have the
same magnitude.

The paper is organized as follows. In Sect. II we cast the
eigenmodes of the Hamiltonian Eq. (\ref{Hamiltonian}) in a convenient notation.
In Sect. III we include recombination
and study its effect on the eigenmodes. The consequences of
nontrivial dynamics for OMAR are considered in Sects. IV and V,
where we perform averaging over realizations of hyperfine fields.
We establish that inequivalence of rms hyperfine fields for
electrons and holes has a dramatic effect on OMAR, when it is
governed by soft pairs. Sect. VI concludes the paper.

\section{Dynamics of a pair in the presence of recombination}
\subsection{Isolated pair}
We start with reviewing the dynamics of a pair of spins in
the absence of recombination. Obviously, this dynamics does
not depend on the choice of the quantization axis. However,
since we plan to include recombination, the choice
of the quantization axis, $z$, illustrated in Fig. \ref{figCoords}
appears to be preferential. The axis is chosen to lie in the
plane containing the vectors ${\bm \Omega}_1$, ${\bm \Omega}_2$.
Moreover, the orientation of the $z$-axis is fixed by the
condition ${\bm \Omega}_{1x}=-{\bm \Omega}_{2x}$. Then the angles,
$\theta_1$, $\theta_2$, between ${\bm \Omega}_1$, ${\bm \Omega}_2$ and
the $z$-axis are given by
\begin{equation}
\tan \theta_1=\frac{|{\bm \Omega}_1 \times {\bm \Omega}_2|}{{\bm \Omega}_2^2+
{\bm \Omega}_1\cdot{\bm \Omega}_2},~~~  
\tan \theta_2=\frac{|{\bm \Omega}_1 \times {\bm \Omega}_2|}{{\bm \Omega}_1^2+
{\bm \Omega}_1\cdot{\bm \Omega}_2}
\end{equation}
With this choice,  the Schr{\"o}dinger equation for the
amplitudes of $S$, $T_0$, $T_{+}$, and $T_{-}$ reduces
to the system
\begin{align}
\label{system}
i \frac{\partial T_{+}}{\partial t} &= \Sigma_{z} T_{+} - \frac{1}{\sqrt{2}} \Delta_{x} S, \\
i \frac{\partial S}{\partial t} &= \Delta_{z} T_0 - \frac{1}{\sqrt{2}} \Delta_{x} T_{+}
+ \frac{1}{\sqrt{2}} \Delta_{x} T_{-}, \\
i \frac{\partial T_{0}}{\partial t} &= \Delta_{z} S, \\
\label{system-end}
i \frac{\partial T_{-}}{\partial t} &= -\Sigma_{z} T_{-} + \frac{1}{\sqrt{2}} \Delta_{x}S,
\end{align}
where $\Sigma_z, \Delta_z,$ and $\Delta_x$ are defined as
\begin{align}
\label{definitions}
\Sigma_z &= \frac{\Omega_{1z} + \Omega_{2z}}{2}, \\
\Delta_{z} &= \frac{\Omega_{1z} - \Omega_{2z}}{2}, \\
\Delta_{x} &= \frac{\Omega_{1x} - \Omega_{2x}}{2}.
\end{align}
The advantage of our choice of the quantization axis shows in the
fact that the state $T_0$ is coupled exclusively to $S$. Since recombination is allowed only from $S$, this will simplify the subsequent analysis of the
recombination dynamics.

The eigenvalues, $\lambda_i$, of the system Eqs. (\ref{system})-(\ref{system-end}) satisfy the quartic equation
\begin{equation}
\label{eqLambda1}
\lambda_i^2(\lambda_i^2 - \Sigma_{z}^2) - \lambda_i^2 (\Delta_{z}^2 + \Delta_{x}^2) +
\Delta_z^2 \Sigma_z^2  = 0.
\end{equation}
We will enumerate these eigenvalues according to the convention $\lambda_1=-\lambda_2$ and $\lambda_3=-\lambda_4$. To find the absolute
values  $\lambda_1^2$, $\lambda_3^2$ one does not have to solve
Eq. (\ref{eqLambda1}), since it is obvious that for non-interacting spins
the eigenvalues are the sums and the differences of individual Zeeman
energies
\begin{equation}
\label {eqEigenvals}
\lambda_1^2=\left( \frac{|{\bm \Omega}_1| + |{\bm \Omega}_2|}{2} \right)^2,
~~\lambda_3^2=\left( \frac{|{\bm \Omega}_1| - |{\bm \Omega}_2|}{2} \right)^2.
\end{equation}
Naturally, $\lambda_1$, $\lambda_3$ do not depend on the choice of the quantization axis. At the same time, the coefficients in Eq. (\ref{eqLambda1})
do depend on this choice. To trace how the dependence on the quantization axis
disappears in the roots of Eq. (\ref{eqLambda1}), one should use the following
identities
\begin{align}
\label{identities}
\Sigma_z^2+\Delta_z^2+\Delta_x^2&=\frac{|{\bm \Omega}_1|^2+|{\bm \Omega}_2|^2}{2},\\
\Sigma_z^2\Delta_z^2&=\left( \frac{|{\bm \Omega}_1|^2-|{\bm \Omega}_2|^2}{4} \right)^2.
\end{align}

In terms of the angles $\theta_1$ and $\theta_2$, Fig. \ref{figCoords}, the corresponding eigenvectors
can be expressed as
\begin{multline}
\begin{pmatrix}
T_{+} \\ S \\ T_{0} \\ T_{-}
\end{pmatrix} =
\left\{
\begin{pmatrix}
    \cos \frac{\theta_1}{2} \cos \frac{\theta{2}}{2} \\
   -\frac{1}{\sqrt{2}} \sin \frac{\theta_1 + \theta_2}{2} \\
    \frac{1}{\sqrt{2}} \sin \frac{\theta_1 - \theta_2}{2} \\
   -\sin\frac{\theta_1}{2}\sin\frac{\theta_2}{2}
\end{pmatrix},
\begin{pmatrix}
  -\sin\frac{\theta_1}{2}\sin\frac{\theta_2}{2}\\
  -\frac{1}{\sqrt{2}} \sin \frac{\theta_1 + \theta_2}{2} \\
  -\frac{1}{\sqrt{2}} \sin \frac{\theta_1 - \theta_2}{2} \\
   \cos \frac{\theta_1}{2} \cos \frac{\theta{2}}{2}
\end{pmatrix}, \right. \\
\left.
\begin{pmatrix}
   \cos \frac{\theta_1}{2} \sin \frac{\theta{2}}{2} \\
   \frac{1}{\sqrt{2}} \cos \frac{\theta_1 + \theta_2}{2} \\
   \frac{1}{\sqrt{2}} \cos \frac{\theta_1 - \theta_2}{2} \\
   \sin\frac{\theta_1}{2}\cos\frac{\theta_2}{2}
\end{pmatrix},
\begin{pmatrix}
  -\sin \frac{\theta_1}{2} \cos \frac{\theta{2}}{2} \\
  -\frac{1}{\sqrt{2}} \cos \frac{\theta_1 + \theta_2}{2} \\
   \frac{1}{\sqrt{2}} \cos \frac{\theta_1 - \theta_2}{2} \\
  -\cos\frac{\theta_1}{2}\sin\frac{\theta_2}{2}
\end{pmatrix}
 \right\},
\label{eqEigenvects}
\end{multline}
where the first two correspond to $\lambda_{1,2}$ while the last two correspond
to $\lambda_{3,4}$, respectively.

The form Eq. (\ref{eqEigenvects}) allows us to make the following observation.
When the full magnetic fields acting on spins incidentally coincide, we have
$|\vc{\Omega}_1| = |\vc{\Omega}_2|$. Then it follows from  Eq. (\ref{eqEigenvals}) that
$\lambda_3 = \lambda_4 = 0$, so that the two corresponding eigenstates become degenerate.
Under this condition we also have $\theta_1 = \theta_2$. Then the first two
eigenvectors Eq. (\ref{eqEigenvects}) have zeros in the rows corresponding to $T_0$.
Concerning the other two eigenvectors, due to their degeneracy, their sum and difference
are also eigenvectors. The difference  has a zero in the row corresponding to $T_0$,
while the sum consists of the $T_0$ component, exclusively.
Then we conclude that for realizations of hyperfine field for which
$|\vc{\Omega}_1| = |\vc{\Omega}_2|$ the state $T_0$ is {\em completely decoupled}
from the other three states. This fact has important implications for recombination
dynamics, as we will see below.

Including recombination requires the analysis of the full equation for
the density matrix
\begin{equation}
\label{full}
i \dot{\rho} = [\widehat{H}, \rho] - \frac{i}{2 \tau} \left\{ \rho, \ket{S}\bra{S} \right\},
\end{equation}
where $\tau$ is the recombination time. The form of the second term ensures that
recombination takes place only from $S$. The matrix corresponding to Eq. (\ref{full})
is $16\times 16$. The $16$ eigenvalues can be cast in the form $\lambda_i - \lambda_j^{\ast}$,
where $\lambda_i$ and $\lambda_j$ satisfy the equation
\begin{equation}
\label{generalized}
\lambda_i\left( \lambda_i + \frac{i}{\tau}  \right)(\lambda_i^2 - \Sigma_z^2)
- \lambda_i^2 (\Delta_{z}^2 + \Delta_{x}^2) +
\Delta_z^2 \Sigma_z^2 = 0.
\end{equation}
The latter equation expresses the condition that
$\lambda_i$ are the eigenvalues of non-hermitian
operator $\widehat{H}-\frac{i}{\tau}\ket{S}\bra{S}$.
In the limit $\tau \rightarrow \infty$ this equation reduces to  Eq. (\ref{eqEigenvals}).
The dynamics of recombination is governed by the imaginary parts of the roots of
Eq. (\ref {generalized}), i.e. decay is described by the exponents
$\exp\left[-\left( \text{Im}\,\lambda_i + \text{Im}\,\lambda_j\right) t\right]$.
Less trivial is that finite $\tau$ can strongly affect the  real parts of $\lambda_i$.
Physically, the dependence of $\text{Re}\,\lambda_i$ on $\tau$ describes the {\em back-action}
of recombination on the dynamics of beating between different eigenstates.
In the following two subsections this effect will be analyzed in detail
in the two limiting cases.

\begin{figure}[t]
\centering
\begin{tabular}{ccc}
\fbox{\includegraphics[height=28mm, clip]{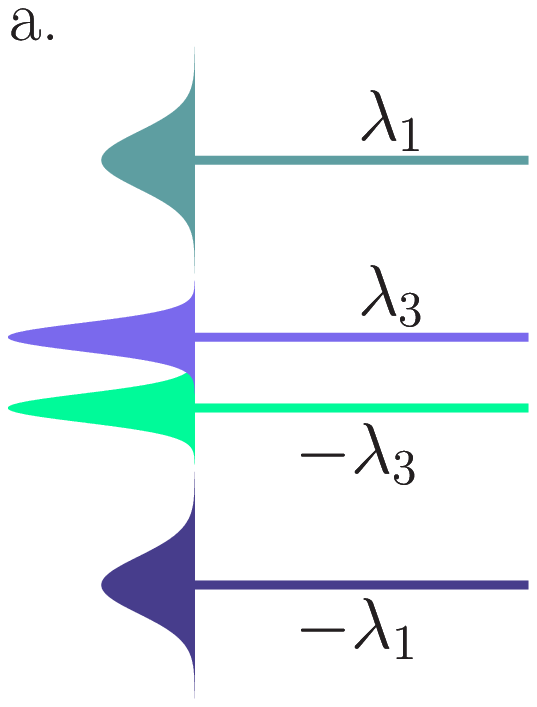}} &
\fbox{\includegraphics[height=28mm, clip]{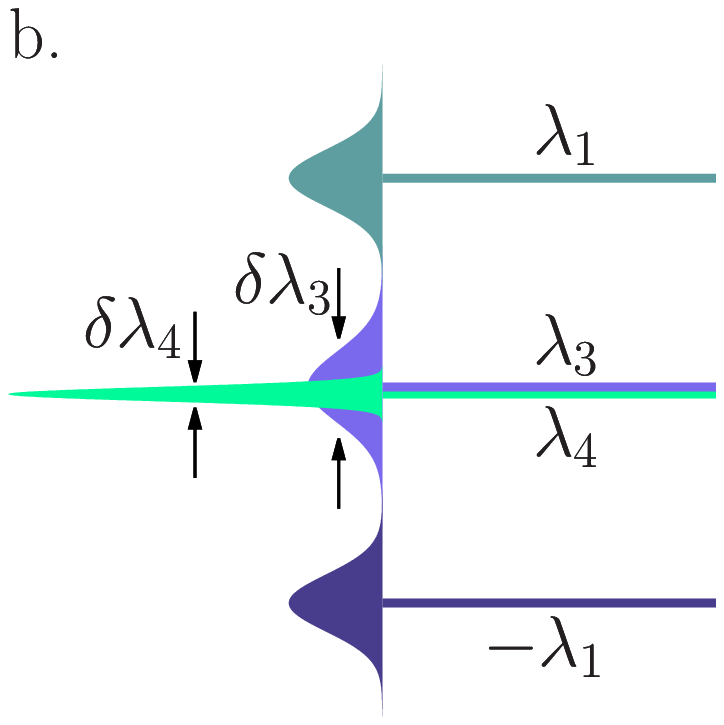}}  &
\fbox{\includegraphics[height=28mm, clip]{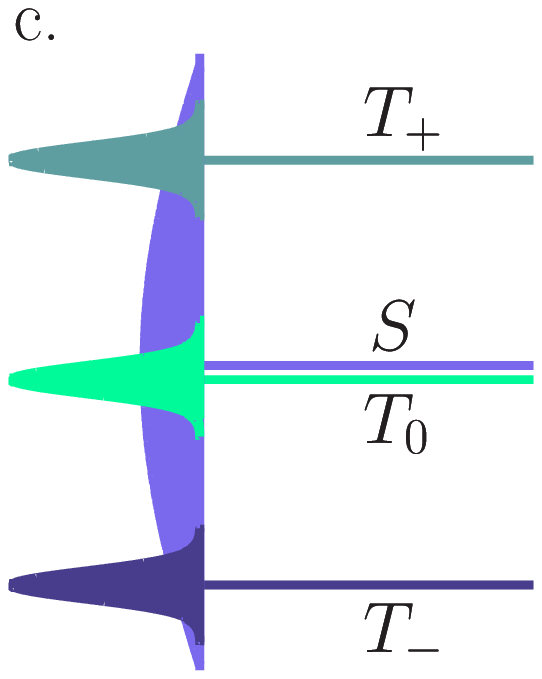}}
\end{tabular}
\caption{(Color online). (a) Slow-recombination regime, $\Omega_{1,2}\ll 1/\tau$. Horizontal lines represent the energy levels Eq. ({\bf 14})  of a pair in the absence of recombination. Recombination
from $S$ causes the broadening of the levels Eqs. ({\bf 21,22}), which, for a typical
pair, is of the same order for all levels. (b) Slow-recombination regime. For {\em soft pairs}, $|{\bm \Omega}_{1}|\approx |{\bm \Omega}_{2}|$, recombination results in splitting Eq. ({\bf 24}) of the {\em widths} of the levels $\lambda_{3,4}$ rather than their positions.
(c) Fast-recombination regime, $\Omega_{1,2}\gg 1/\tau$. The eigenstates $S$, $T_{0}$, $T_{\s +}$, and $T_{\s -}$ are well-defined. Recombination causes strong broadening, $1/\tau$, of the level $S$, and weak broadening $\sim \Omega_{1,2}^2\tau$ of the other
three levels.}
\label{figEnergy}
\end{figure}

\subsection{Slow Recombination}
Consider the limit $1/\tau \ll \lambda_i$. In this limit recombination amounts
to the small corrections
to the bare values of $\lambda_i$ given by Eq. (\ref{eqEigenvals}). This allows one to set
 $\lambda_i$ equal to their bare values in all terms in Eq. (\ref{generalized}) containing
$1/\tau$, and search
for solution in the form $\lambda_i+\delta \lambda_i$.
Then one gets the following expression for the correction $\delta \lambda_i$

\begin{equation}
\delta\lambda_i  = \frac{-i}{\tau}\,\frac{\lambda_i(\lambda_i^2 -\Sigma_z^2)}{
    \lambda_i^3 - 2 \lambda_i (\Sigma_z^2 + \Delta_z^2 + \Delta_x^2)}
  = \frac{-i}{2\tau}\,\frac{\lambda_i^2(\lambda_i^2 - \Sigma_z^2)}{
    \lambda_i^4 - \Sigma_z^2 \Delta_z^2}.
\end{equation}
In the last identity we have used the fact that $\lambda_i$ satisfy the
equation Eq. (\ref{eqEigenvals}). The above expression can be greatly simplified
with the help of the relations Eq. (\ref{identities}). One has
 \begin{align}
\label{eqDeltaLambda}
\delta \lambda_{1,2}
	&= -\frac{i}{4\tau}\left(1-\frac{\vc{\Omega}_1\cdot \vc{\Omega}_2}{
		|{\bm \Omega}_1||{\bm \Omega}_2|}\right),\\
\delta \lambda_{3,4}
	&= -\frac{i}{4\tau}\left(1+\frac{\vc{\Omega}_1\cdot \vc{\Omega}_2}{
		|{\bm \Omega}_1||{\bm \Omega}_2|}\right).
\end{align}
The above result suggests that for generic mutual orientations of ${\bm \Omega}_1$ and ${\bm \Omega}_2$ all modes of a pair decay  with characteristic
time $\sim \tau$. At the same time, for parallel orientations
of  ${\bm \Omega}_1$, ${\bm \Omega}_2$ the modes $\lambda_{1,2}$
have anomalously long lifetime. This long lifetime has its origin
in the fact that for ${\bm \Omega}_1\parallel {\bm \Omega}_2$, the states
$T_{+}$ and $T_{-}$, which are orthogonal to $S$, are the eigenstates of the Hamiltonian Eq. (\ref{Hamiltonian}). Formally this can be seen from the
general expression Eq. (\ref{eqEigenvects}) for the eigenvectors upon setting
$\theta_1=\theta_2=0$. Similarly, for ${\bm \Omega}_1$ and ${\bm \Omega}_2$ being antiparallel, one can check from Eq. (\ref{eqEigenvects}) that for $\theta_1=\pi-\theta_2$, the eigenstates $\lambda_3$, $\lambda_4$ have no
$S$ component, so they are long-lived. Note that  the existence of long lifetimes for parallel and antiparallel
configurations of  ${\bm \Omega}_1$, ${\bm \Omega}_2$ is at the core of
the ``blocking mechanism'' of OMAR proposed in Ref. \onlinecite{Bobbert1}.

\subsubsection{Soft pairs}
As was pointed out in the Introduction, recombination also has
 a pronounced effect on the spin dynamics for sparse configurations
for which $\vert \Omega_1\vert \approx \vert \Omega_2\vert$.
Indeed, for these configurations, the values $\lambda_3$ and $\lambda_4$ are
anomalously small. Then the basic condition, $1/\tau \ll \lambda_i$, under which
Eq. (\ref{eqDeltaLambda}) was derived, is  not satisfied.
We dub such realizations as soft pairs.
For soft pairs the  expressions for $\delta\lambda_1$,
$\delta\lambda_2$ remain valid, but the eigenvalues $\lambda_3$,
$\lambda_4$ get strongly modified due to finite recombination time, $\tau$.

Although for soft pairs  the terms $\propto 1/\tau$
in Eq. (\ref{generalized}) cannot be treated  as a perturbation, a different simplification
becomes possible in this case. We can neglect $\lambda_i^2$
 compared to $\Sigma_z^2$ in the first term and $\Delta_z^2$ compared to $\Delta_x^2$
 in the second term. The first simplification is justified, since the typical value of $\Sigma_z$
is $\sim |{\bm \Omega}_1|\approx |{\bm \Omega}_2|$ and is much bigger than
both $1/\tau$ and $(|{\bm \Omega}_1|-|{\bm \Omega}_2|)$. Concerning the second
simplification, the smallness of  $(|{\bm \Omega}_1|-|{\bm \Omega}_2|)$ automatically implies that $\Delta_z$ given by Eq. (\ref{definitions}) is small. With the above simplifications the eigenvalues $\lambda_{3,4}$ satisfy the quadratic equation
\begin{equation}
\label{quadratic}
(\Sigma_z^2+\Delta_x^2)\lambda_i^2+\frac{i}{\tau}\Sigma_z^2\lambda_i
-\Delta_z^2\Sigma_z^2=0.
\end{equation}
Already from the form of Eq. (\ref{quadratic}) one can make a
surprising observation that, even with finite $1/ \tau$, one of the roots is {\em identically}
zero when $\Delta_z=0$, i.e. when  $|{\bm \Omega}_1|$ and $|{\bm \Omega}_2|$
are exactly equal to each other. This suggests that a pair in the
state corresponding to this root will never recombine. For a small
but finite difference $(|{\bm \Omega}_1|-|{\bm \Omega}_2|)$ the recombination
will eventually take place but {\em only after time much longer than} $\tau$.
Indeed, for the generic case, $|{\bm \Omega}_1| \sim |{\bm \Omega}_2|)$, we have from Eq. (\ref{quadratic})
\begin{equation}
\label{lambda34}
\lambda_{3,4}=-\frac{i}{2\tau}\Bigl[\Lambda\pm \sqrt{\Lambda^2-4\Lambda\Delta_z^2\tau^2}\Bigr],
\end{equation}
where the dimensionless parameter $\Lambda$ is defined as
\begin{equation}
\label{Lambda}
\Lambda=\frac{\Sigma_z^2}{\Sigma_z^2+\Delta_x^2}.
\end{equation}
Even when $|{\bm \Omega}_1|$ and $|{\bm \Omega}_2|$
are close, a typical value of parameter $\Lambda$ is $\sim 1$.  Then Eq. (\ref{lambda34}) suggests that anomalously long-living mode exists in the domain
$\Delta_z \lesssim 1/\tau$ where its lifetime is $\sim 1/\Delta_z^2\tau$. Note that the lifetime becomes {\em longer}
with a {\em decrease} of the recombination time.

As the difference
$|{\bm \Omega}_1|-|{\bm \Omega}_2|$ increases, the product $\Delta_z\tau$ becomes big and the expression under the square root
in Eq. (\ref{lambda34}) becomes negative. Then the lifetimes of
of both states corresponding to $\lambda_3$ and $\lambda_4$ become
equal to $\tau/\Lambda$. Note that, at the same time, the splitting of the real parts of $\lambda_3$ and $\lambda_4$ becomes $\sim \Delta_z^2\tau$, which is {\em much bigger} than $|{\bm \Omega}_1|-|{\bm \Omega}_2|$.

The above effect can be interpreted as a {\em repulsion}
of the eigenvalues caused by recombination\cite{Gefen}.
A more prominent analogy
can be found in optics\cite{Dicke}. The signs $+$ and $-$ in
Eq. (\ref{lambda34}) can be related to the superradiant and subradiant modes of two identical emitters. The role of $\tau$ in this case is played by their radiative lifetime.

 Both effects illustrate the back-action of recombination on the dynamics of the pair when the spin levels of pair-partners are
nearly degenerate. To track an analogy to this effect one can
refer to Refs. \onlinecite{subradiance1} and \onlinecite{subradiance2}, where
Eq. (\ref{lambda34}) appeared in connection to resonant tunneling
through a pair of nearly degenerate levels, while the role of $1/\tau$ was played by the level width with respect to escape into the leads.

For our choice of the quantization axis the long-living state corresponds 
to $T_0$.  For completeness we rewrite the parameter $\Delta_z$,
which enters Eq. (\ref{lambda34}), in the coordinate-independent form 
\begin{equation}
\label{DeltaZ}
\Delta_z^2 = \frac{(|{\bm \Omega}_1|^2 - |{\bm \Omega}_2|^2)^2 }{
4|{\bm \Omega}_1 + {\bm \Omega}_2|^2}.
\end{equation}
To establish coordinate-independent form of parameter $\Lambda$ we
need the combinations $\Sigma_z^2$ and $\Sigma_z^2+\Delta_x^2$, which
are given by
\begin{align}
\label{SigmaZ}
\Sigma_z^2 &= \frac{|{\bm \Omega}_1 + {\bm \Omega}_2|^2}{4}, \\                     \Sigma_z^2+\Delta_x^2 &= \frac{|{\bm \Omega}_1|^2 + |{\bm \Omega}_2|^2}{2}
- \frac{
      \left( |{\bm \Omega}_1|^2 - |{\bm \Omega}_2|^2\right)^2}
    {4 |{\bm \Omega}_1 + {\bm \Omega}_2|^2 } ,
\end{align}
so that $\Lambda$ can be cast into the form
\begin{equation}
\label{LambdaVect}
\Lambda=\frac{|{\bm \Omega}_1+{\bm \Omega}_2|^4}
{|{\bm \Omega}_1+{\bm \Omega}_2|^4+4|{\bm \Omega}_1\times{\bm \Omega}_2|^2}.
\end{equation}

The consequences of ``trapping" described by Eq. (\ref{lambda34})
for OMAR will be considered in Sections IV and V. In the subsequent
subsection we will see that the similar physics, namely, the emergence of {\em slow} modes due to {\em fast} recombination persists also
in the domain $|{\bm \Omega}_{1,2}|\tau \ll 1$.

\subsection{Fast Recombination}
In the opposite limit, $\tau \ll |{\bm \Omega}_{1,2}|^{-1}$, the bracket $(\lambda_i +\frac{i}{\tau})$
in  Eq. (\ref{generalized}) is big. This suggests that three zero-order eigenvalues
are
\begin{equation}
\label{firstthree}
\lambda_i = 0, \pm \Sigma_z.
\end{equation}
In the same order, the fourth eigenvalue is $-\frac{i}{\tau}$.
Concerning the eigenvectors, in the zeroth order they are simply
$S, T_{+}, T_{-},$ and $T_0$.
 This follows from the equation
\begin{equation}
i \dot{S} + \frac{i}{\tau}S = \Delta_z T_0  - \frac{1}{\sqrt{2}} T_{+}
+ \frac{1}{\sqrt{2}} T_{-}.
\end{equation}
Taking $\tau$ to zero means that in the zeroth order $S=0$. Then
three other equations in the system Eq. (\ref{system}) get decoupled.

In the first order, the eigenvalues Eq. (\ref{firstthree}) acquire imaginary
parts
\begin{equation}
\label{imaginaryparts}
\delta\lambda_i= - i\tau\left( \frac{\lambda_i^2(\Delta_z^2 + \Delta_x^2)
    -\Delta_z^2\Sigma_z^2}{3 \lambda_i^2 - \Sigma_z^2}
\right).
\end{equation}
With the help of Eqs. (\ref{SigmaZ}) and (\ref{LambdaVect}) these imaginary parts
can be simplified to
\begin{align}
\label{imaginary1}
\delta \lambda_{T_0}
	&= -i \tau \Delta_z^2 = -i \tau \frac{(\Omega_1^2 -\Omega_2^2)^2}{
	4|\vc{\Omega}_1 + \vc{\Omega}_2|^2}, \\
\label{imaginary2}
\delta \lambda_{T_{+}} = \delta \lambda_{T_{-}}
	&= -i \tau \frac{\Delta_x^2}{2} =  - i \tau \frac{|\vc{\Omega}_1 \times \vc{\Omega}_2|^2}
	{2|\vc{\Omega}_1 + \vc{\Omega}_2|^2}.
\end{align}
We see that for a generic situation
$|\vc{\Omega}_1|\sim |\vc{\Omega}_2|$ the lifetime of
the modes $T_0$, $T_{+}$, and $T_{-}$ are $\sim 1/|\vc{\Omega}_{1,2}|^2\tau$, i.e. in the regime of fast recombination it is much longer than
$\tau$. This is a consequence of effective decoupling of $T_0$, $T_{+}$, and $T_{-}$ from $S$ in this regime. We also observe from
Eq. (\ref{imaginary1}) that there is additional prolongation
of lifetime for the mode $T_0$ if the pair is soft.
Eq. (\ref{imaginary1}) also suggests
that lifetimes of the states $T_{+}$, $T_{-}$ are anomalously long when $\vc{\Omega}_1$ and $\vc{\Omega}_2$ are collinear. This expresses the obvious
fact that, for {\em collinear} effective fields acting on the
pair-partners, $T_{+}$ and $T_{-}$ are the eigenstates
no matter whether recombination is present or not.

Once the eigenvalues and eigenvectors of a pair in
the presence of recombination are established, the
next question crucial for transport through the pair
is: Suppose that initial state is a random superposition
of $S$, $T_0$, $T_{+}$, and $T_{-}$, what is the average (over the coefficients of superposition) waiting time
for this state to recombine? Naturally, the answer to this question does not depend on the actual choice of the orthonormal basis. We address this question
in the next section.

\section{Recombination time from a random initial state}

\subsection{Soft pair in a slow recombination regime}
To illustrate the peculiarity of the question posed above,
we start from an instructive particular case of soft pair in a slow recombination regime. We defined a soft pair
as a pair for which the condition
$(|{\bm \Omega}_1| - |{\bm \Omega}_2|)\ll |{\bm \Omega}_{1,2}|$ is met. However, in the slow recombination regime, the combination $(|{\bm \Omega}_1| - |{\bm \Omega}_2|)\tau$ can be either big or small.
In both cases there is a strong separation between the absolute values of
$\lambda_{1,2}$ and $\lambda_{3,4}$.
It can be seen from Eq. (\ref{lambda34}) that in the limit $(|{\bm \Omega}_1| -|{\bm \Omega}_2|)\tau \gg 1$,
the recombination times for states which correspond to
$\lambda_3$ and $\lambda_4$ are given by
\begin{align}
\label{tLarge}
t_{\s R}^{\s (3)} = \frac{2\tau}{\Lambda}~~~\text{and}~~~
t_{\s R}^{\s (4)} = \frac{2\tau}{\Lambda},
\end{align}
 while in the opposite limit, $(|{\bm \Omega}_1| -|{\bm \Omega}_2|)\tau \ll 1$, we get
\begin{equation}
\label{tSmall}
t_{\s R}^{\s (3)}  = \frac{2\tau}{\Lambda}~~~\text{and}~~~
t_{\s R}^{\s (4)} = \frac{1}{\tau \Delta_z^2}.
\end{equation}
We see that the recombination time of $\lambda_3$ is $\sim \tau$ for both limits, while the recombination time of
$\lambda_4$ crosses over from $\sim \tau$ to $\sim 1/\Delta_z^2\tau$ as   $(|{\bm \Omega}_1| -|{\bm \Omega}_2|)\tau$ decreases. Taking into account that
for generic case $|{\bm \Omega}_1| \sim |{\bm \Omega}_2|$
the recombination times corresponding to $\lambda_{1,2}$
are $\sim \tau$, we conclude that for purely random initial conditions the average recombination time is either $\sim \tau$ or it is $\frac{1}{4}$ of
$1/\Delta_{z}^2\tau$.

The major complication for getting exact average
recombination time for a soft pair is that the
exact eigenstates represent mixtures with
weights governed by the recombination time. This
follows from Eq. (\ref{lambda34}). In addition, the
eigenstates corresponding to $\lambda_{3}$, and $\lambda_4$
are not orthogonal to each other.
However, for a soft pair these complications
can be overcome. The reason is that, there are two small
parameters in the problem, $1/\tau |{\bm \Omega}_{1,2}|$,
and $(|{\bm \Omega}_{1}|-|{\bm \Omega}_{2}|)/|{\bm \Omega}_{1,2}|$.
The first parameter guarantees slow recombination, while
the second ensures that the pair is soft.
The presence of these parameters allows us to evaluate $\langle t_{\s R} \rangle$
in the closed form using the general formula
\begin{equation}
\label{General}
\langle t_{\s R} \rangle = \frac{1}{4} \sum_{i,j} g_{ji}(g^{-1}_{ji})^{*} \frac{1}{i(\lambda_i - \lambda_j^{*})},
\end{equation}
where  $g_{ij} = \left<v_i|v_j\right>$ is a matrix of inner products
of eigenvectors corresponding to {\em complex} eigenvalues $\lambda_i$ and
$\lambda_j$. The above formula becomes absolutely transparent
when the eigenvectors are orthonormal. Then the matrix $g_{ij}$ reduces
to the Kronecker symbol, $\delta_{ij}$, and $\langle t_{\s R} \rangle$ simplifies
to
\begin{equation}
\label{Particular}
\langle t_{\s R} \rangle
= - \frac{1}{8} \sum_j \frac{1}{\text{Im}\,\lambda_j},
\end{equation}
which expresses the fact that for random initial state the
average recombination time is the evenly-weighted sum of
recombination times from eigenstates.

In the case of a soft pair and slow recombination one should
use Eq. (\ref{General}) to evaluate $\langle t_{\s R} \rangle$.
What enables this evaluation is that, by virtue of small parameters,
the eigenvectors corresponding to $\lambda_1$ and $\lambda_2$ are
mutually orthogonal (with accuracy $1/\tau |{\bm \Omega}_{1,2}|$), and they
are both orthogonal to eigenvectors corresponding to $\lambda_3$ and
$\lambda_4$. Therefore, in evaluating Eq. (\ref{General}), one has to deal
only with mutual non-orthogonality of two eigenvectors $v_3$ and $v_4$.
The straightforward calculation yields
\begin{equation}
\label{result}
\langle t_{\s R} \rangle = \frac{\tau}{\Lambda}
	+ \frac{1}{4\Delta_z^2 \tau}-\frac{1}{\text{Im}\,\lambda_1}
		 -\frac{1}{\text{Im}\,\lambda_2},
\end{equation}
where ${\text{Im}\,\lambda_1}={\text{Im}\,\lambda_2}$ are given by Eq. (\ref{eqDeltaLambda}).
It is easy to see that in the limiting cases of large and small
$(|{\bm \Omega}_1| -|{\bm \Omega}_2|)\tau$ Eq. (\ref{result}) reproduces
Eqs. (\ref{tLarge}) and (\ref{tSmall}), respectively.

\begin{figure}[t]
\includegraphics[width=77mm, clip]{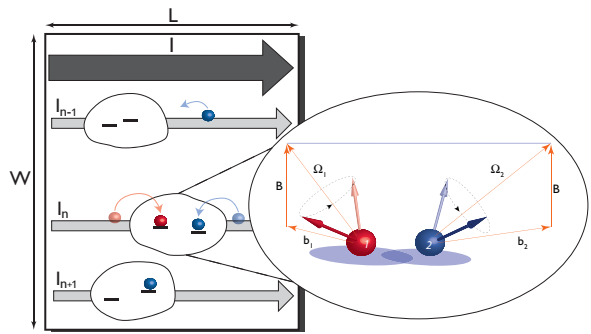}
\caption{(Color online). The simplest  model of transport  through a bipolar device in which the currents flow
along independent chains.
Electrons arrive at the recombination region from the left, while the holes arrive from the right.
Blobs enclose the sites from which electron and hole recombine. One of the blobs is
enlarged to illustrate  the spin precession of the pair
partners in their respective fields ${\bm \Omega}_1$, ${\bm \Omega}_2$. For {\em soft} pairs
the magnitudes of ${\bm \Omega}_1$ and ${\bm \Omega}_2$ are close to each other.}
\label{figModel}
\end{figure}

While in the last two terms in Eq. (\ref{result}) depend weakly
on the degree of ``softness" of the pair, $\Delta_z\propto (|{\bm \Omega}_{1}|-|{\bm \Omega}_{2}|)$, the second term exhibits {\em unlimited} growth with decreasing $\Delta_z$. We emphasize the peculiarity of this situation. In conventional quantum mechanics, when the level separation becomes smaller than their width, it should be simply replaced by the width. What makes Eq. (\ref{result}) special is that the smaller is
$\Delta_z$ the more the state $T_0$ becomes isolated. There is direct analogy
of this situation with the Dicke effect\cite{Dicke}, as was mentioned in the Introduction.
By virtue of this analogy, the state $T_0$ assumes the role of the ``subradiant"
mode which accompanies the formation of the superradiant mode. In the Dicke effect
the formation of superradiant and subradiant states occurs because the bare states are
coupled via continuum. In our situation it is recombination that is responsible
for ``isolation" of  $T_{ 0}$.
If the pair is not soft, the calculation of the time $\langle t_{\s R}\rangle$
in the slow-hopping regime can be performed by simply using Eq. (\ref{Particular})
and $\lambda_{i}$ given by Eqs. (\ref{eqDeltaLambda}), (\ref{lambda34}). This is because
the smallness of $1/\tau$ makes the eigenstates almost orthogonal.
However, the Dicke physics becomes even more pronounced in the fast-recombination regime, as
demonstrated in the next subsection.

\subsection{Recombination time in the fast recombination regime}
It might seem that under the condition of fast recombination
$|{\bm \Omega}_{1,2}|\tau \ll 1$ the recombination time from
the random initial state should be $\sim \tau$, since
spins practically do not precess during the time $\tau$.
The fact that recombination takes place only from $S$,
while initial state is  a random mixture, already suggests
that $\langle t_{\s R}\rangle$ is longer than $\tau$.
This is because if the initial configuration  is different from $S$
it must first cross over into $S$ by spin precession
before it recombines. The characteristic time for the
spin precession is $\sim |{\bm \Omega}_{1,2}|^{-1} \gg \tau$.
It turns out that the crossing time is actually much longer
than $|{\bm \Omega}_{1,2}|^{-1}$. Formally, this fact follows from Eqs. (\ref{imaginary1}), (\ref{imaginary2}) for $\delta \lambda_i$, which are
of the order of $|{\bm \Omega}_{1,2}|^2\tau$ rather than $|{\bm \Omega}_{1,2}|$. We can now interpret this result by identifying
$S$ with superradiant state, while $T_{0}$, $T_{+}$, and $T_{-}$
assume the roles of subradiant states. The short lifetime of  $S$ isolates
it from the rest of the system. Quantitatively, the portion of
$S$ in the other eigenvectors is  $\sim |{\bm \Omega}_{1,2}|\tau$.

What is important for calculation
of $\langle t_{\s R}\rangle$ is the fact that eigenvectors are
orthogonal (with accuracy $\sim 1/|{\bm \Omega}_{1,2}|\tau$) in the
fast-recombination regime. This allows one to replace the overlap integrals $g_{\s ij}$
in Eq. (\ref{General}) by $\delta_{\s ij}$ and use the Eq. (\ref{Particular})
which immediately yields for  $\langle t_{\s R}\rangle$
the result
\begin{align}
\langle t_{\s R} \rangle &= - \frac{1}{8} \left( \frac{1}{\text{Im}\,\lambda_S}
+ \frac{1}{\text{Im}\,\lambda_{T_0}}
+ \frac{1}{\text{Im}\,\lambda_{T_{+}}}
+ \frac{1}{\text{Im}\,\lambda_{T_{-}}}
\right) \\
 &= \frac{1}{8} \left[ \tau + \frac{1}{\tau}\left(
 \frac{1}{\Delta_z^2} + \frac{4}{\Delta_x^2}
 \right) \right].
\end{align}
Substituting the coordinate-independent expressions for $\Delta_{x}$ and
$\Delta_{z}$, we arrive at the final expression for recombination time,
which is applicable within the entire fast-recombination regime
\begin{equation}
\label{tR-fast}
\langle t_{\s R} \rangle =
\frac{1}{8} \left[
\tau + \frac{4}{\tau} \left(
\frac{|{\bm \Omega}_1 + {\bm \Omega}_2 |^2}{
    \left( |{\bm \Omega}_1|^2 - |{\bm \Omega}_2|^2 \right)^2}
  + \frac{|{\bm \Omega}_1 + {\bm \Omega}_2 |^2}{
    | {\bm \Omega}_1 \times {\bm \Omega}_2 |^2}
\right)
\right].
\end{equation}
As was already noticed in the previous section, recombination time
diverges for two particular configurations: soft pairs with
$|{\bm \Omega}_1| =|{\bm \Omega}_2|$ and collinear ${\bm \Omega}_1$
and ${\bm \Omega}_2$. Certainly this divergence will be cut off in
the course of calculation of current through a pair to which we now turn.

\begin{figure}[t]
\includegraphics[width=77mm, clip]{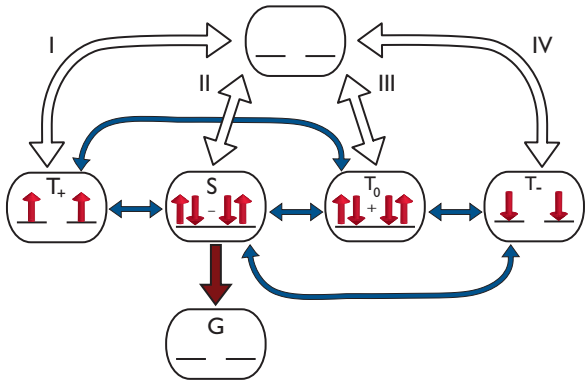}
\caption{(Color online). I, II, III, and IV are possible variants of the current cycle. For each variant the pair is initially  created in one of four states.
This is followed by time evolution, illustrated by blue double arrows, which mixes the
states. Subsequently, the pair either recombines from $S$ (brown arrow) or  dissociates. The processes of creation and dissociation are indicated by white double arrows.
The current is the inverse duration, $\overline{t}$, of the cycle  averaged over  initial states, which we assume to have equal probabilities. The time, $\overline{t}$, is  given by Eqs. ({\bf 44}),  ({\bf 45}), or ({\bf 46}) depending on the recombination regime.}
\label{figCycle}
\end{figure}

\section{Transport model}

We adopt a transport model illustrated in Fig. \ref{figModel}.
For concreteness we will discuss a bipolar device, so that
the current is due to electron-hole recombination.
As shown in Fig. \ref{figModel}, electrons arrive at
the pair of sites (enlarged regions in Fig. \ref{figModel})
from the left, while holes arrive from
the right. Once an electron-hole pair is formed, the spins of
the pair-partners undergo precession in the fields ${\bm \Omega}_1$
and ${\bm \Omega}_2$, respectively, waiting to either recombine or
to bypass each other and proceed along their respective current paths. For simplicity
we choose the current paths in the form of $1D$ chains. This choice
makes the adopted model of transport
very close to the ``two-site" model proposed in Ref. \onlinecite{Bobbert1}.
The on-site dynamics of a pair {\em with recombination} was  studied in
detail in previous sections. To utilize the results of Sect. III
for the calculation of current, $I$, one has to  incorporate the stages
of formation and dissociation of pairs into the description of transport.

In Fig. \ref{figCycle} the formation and dissociation
are illustrated with white double-sided arrows. The
formation time for all four variants of initial
states is assumed to be the same,  $\tau_{\s D}$.
For simplicity we choose the average time for bypassing
to be also $\tau_{\s D}$. Note that this choice does not limit the generality
of the description, provided that $\tau_{\s D}$ is longer than the recombination time.
The middle and the bottom portions in Fig. \ref{figCycle} illustrate the  spin precession (blue arrows) and recombination (brown arrow) stages, which we
studied earlier. Implicit in  Fig. \ref{figCycle},
is that the pair disappears either due to dissociation or by recombination before the next charge carrier arrives.
Another way to express this fact
is to state that the passage of current proceeds in {\em cycles}.

Naturally, subsequent cycles are statistically independent. This allows
one to express the current along a $1D$ path through the average duration
of the cycle, $\overline{t}$. Indeed,  $N\gg 1$
 cycles take the time $T_{\s N}=t_1 + t_2 + \cdots + t_{\s n}$.
For large $N$, this net time acquires a gaussian distribution
centered at $\overline{T_{\s N}}=N\overline{t}$. Correspondingly,
the current, $N/T_{\s N}$, saturates at the  value
\begin{equation}
\label{current}
I=\frac{1}{\overline{t}}.
\end{equation}
Note, that Eq. (\ref{current}) constitutes an alternative approach
to solving the system of rate equations for two-site model, as in
Ref. \onlinecite{Bobbert1}, or to solving numerically the steady-state
density-matrix equations, as in Ref. \onlinecite{Bobbert2}. Note also,
that Eq. (\ref{current}) is applicable to such singular
realizations as soft pairs, while  previous approaches are not.
For detailed discussion of this delicate point see
Ref. \onlinecite{subradiance2}.

The remaining task is to express $\overline{t}$ via the average recombination
time, $\langle t_{\s R} \rangle$ and $\tau_{\s D}$.
For a typical pair in the regime of slow recombination  $\langle t_{\s R} \rangle$
is given by Eq. (\ref{Particular}) upon substitution of Eq. (\ref{eqDeltaLambda}).
Using this expression we get for
average duration of the cycle
\begin{widetext}
\begin{equation}
\label{duration1}
\overline{t} = \tau_{\s D} +
\frac{1}{4} \left[2\!\times\!
\frac{1}{\frac{1}{\tau}\left(1 - \frac{\vc{\Omega}_1 \cdot \vc{\Omega}_2}{
    |\vc{\Omega}_1||\vc{\Omega}_2|} \right) + \frac{1}{\tau_{\s D}}}
+ 2\!\times\!\frac{1}{\frac{1}{\tau}\left( 1 + \frac{\vc{\Omega}_1 \cdot \vc{\Omega}_2}{
	|\vc{\Omega}_1| |\vc{\Omega}_2|} \right) + \frac{1}{\tau_{\s D}}}
\right].
\end{equation}
\end{widetext}
The first term captures the formation of the pair, while $1/\tau_{\s D}$ in the
denominators describes the bypassing. Indeed, if recombination times are $\sim \tau$,
one can neglect $1/\tau_{\s D}$ in the denominators.
On the other hand, as the brackets in denominators
in Eq. (\ref{duration1}) turn to zero, which corresponds to anomalously slow recombination, the second term becomes $\tau_{\s D}$.
Similarly, for slow recombination  with soft pairs, using Eq. (\ref{result}) we get
\begin{widetext}
\begin{equation}
\label{duration2}
\overline{t} = \tau_{\s D} +
\frac{1}{4} \left[
2\!\times\!\frac{1}{\frac{1}{\tau}\left(1 - \frac{\vc{\Omega}_1 \cdot \vc{\Omega}_2}{
	|\vc{\Omega}_1| |\vc{\Omega}_2|} \right) + \frac{1}{\tau_{\s D}}} 
+ \frac{1}{\frac{1}{\tau}\left( \frac{|\vc{\Omega}_1 + \vc{\Omega}_2|^4}{
	|\vc{\Omega}_1 + \vc{\Omega}_2|^4
		+ 4 | \vc{\Omega}_1 \times \vc{\Omega}_2|^2} \right)
		+ \frac{1}{\tau_{\s D}}}
+ \frac{1}{ \tau \left( \frac{(|\vc{\Omega}_1|^2 - |\vc{\Omega}_2|^2)^2}{
	|\vc{\Omega}_1 + \vc{\Omega}_2|^2} \right) + \frac{1}{\tau_{\s D}} }
\right].
\end{equation}
\end{widetext}
Finally, in the regime of fast recombination one should use Eq. (\ref{tR-fast}) for $\langle t_{\s R} \rangle$. This leads to the following expression for $\overline{t}$
\begin{widetext}
\begin{equation}
\label{duration3}
\overline{t} = \tau_{\s D} +
\frac{1}{4} \left[
\frac{1}{ \frac{1}{\tau} + \frac{1}{\tau_{\s D}} }
+ \frac{1}{  \tau\left( \frac{(|\vc{\Omega}_1|^2 - |\vc{\Omega}_2|^2)^2}{
	4|\vc{\Omega}_1 + \vc{\Omega}_2|^2} \right) + \frac{1}{\tau_{\s D}}}
+ 2\!\times\!\frac{1}{  \frac{\tau}{2}\left( \frac{|\vc{\Omega}_1 \times \vc{\Omega}_2|^2}{
	|\vc{\Omega}_1 + \vc{\Omega}_2|^2} \right) + \frac{1}{\tau_{\s D}}}
\right].
\end{equation}
\end{widetext}

Obviously, the dependence of current on external field is encoded in Eqs. (\ref{duration1})-(\ref{duration3}) via the frequencies ${\bm \Omega}_1={\bm B}+ {\bm b}_e$ and ${\bm \Omega}_2={\bm B}+ {\bm b}_h$. The observable is the current averaged
over realizations of the hyperfine fields ${\bm b}_e$ and ${\bm b}_h$. This averaging
is performed in the next section.

\section{Averaging over hyperfine fields}

\subsection{Averaging in the  slow-recombination regime }

\begin{figure}[t]
\includegraphics[width=77mm, clip]{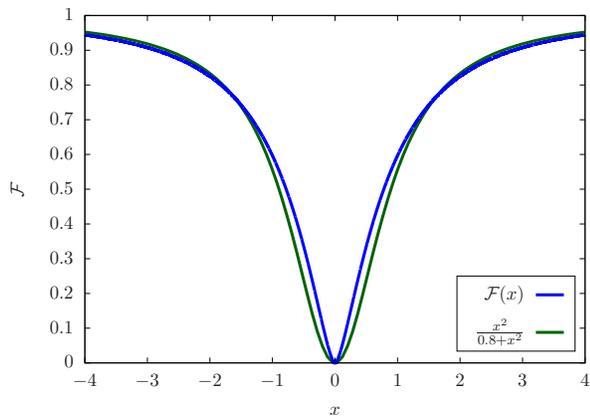}
\caption{(Color online). Blue line: Magnetic field response, $\delta I_{t}(B)$, for the ``parallel-antiparallel"  blocking mechanism,
 is plotted from Eq. (\ref{E1}) in the units $1/\tau_{\s D}$ versus dimensionless
magnetic field $B/B_{\s c}$. Green line: fit with conventional lineshape of OMAR,
$x^2/(0.8+x^2)$. }
\label{figTypical}
\end{figure}

Our basic assumption is that the time, $\tau_{\s D}$,
of formation and dissociation of a pair is much bigger
than the recombination time, $\tau$. Only under this
condition the pair will exercise the spin dynamics.
Using the relation $\tau_{\s D}\gg \tau$, we can simplify
the expression Eq. (\ref{duration1}) for  ${\overline t}$ of
a {\em typical} pair
\begin{equation}
\overline{t} = \tau_{\s D} + \frac{\tau}{
	1 - \left( \frac{\vc{\Omega}_1 \cdot \vc{\Omega}_2}{
		|\vc{\Omega}_1| |\vc{\Omega}_2|} \right)^2}.
\end{equation}
We can also rewrite the current in the form $I=\frac{1}{\tau_{\s D}}-\delta I_{t}(B)$,
where the field-dependent correction is defined as

\begin{equation}
\label{deltaI}
\delta I_{t}(B) =   \frac{\tau}{\tau_{\s D}^2}
	\frac{1}{1 - \left( \frac{\vc{\Omega}_1 \cdot \vc{\Omega}_2}{
		|\vc{\Omega}_1| |\vc{\Omega}_2|}\right)^2 + \frac{\tau}{\tau_{\s D}} }
\end{equation}
As we will see below, the significant change of $\delta I_{t}$ with $B$ takes
place in the domain where $B$ is much bigger than the hyperfine field.
Therefore,  we expand Eq. (\ref{deltaI}) with respect to $|{\bm b}_{\s e}|/B$
and $|{\bm b}_{\s h}|/B$. The principal ingredient of this step is the expansion
of denominator
\begin{multline}
\label{expansion}
	|\vc{\Omega}_1|^2 |\vc{\Omega}_2|^2
	- (\vc{\Omega}_1 \cdot \vc{\Omega}_2)^2 \\
	\approx B^2\left[ |\vc{b}_{\s e} - \vc{b}_{\s h}|^2
	- \left(
		\vc{b}_{\s e} \cdot \frac{\vc{B}}{B} -
		\vc{b}_{\s h} \cdot \frac{\vc{B}}{B} \right)^2 \right].
\end{multline}
Assuming identical Gaussian distributions of ${\bm b}_{\s e}$,
${\bm b}_{\s h}$
\begin{equation}
\label{Gaussian}
\mathcal{P}(\vc{b}_i) = \frac{1}{(\pi b_0)^{3/2}}
\exp (- |\vc{b}_i|^2/b_0^2),
\end{equation}
and choosing the $z$-direction along ${\bm B}$ we get
\begin{equation}
\label{deltaIav1}
\langle \delta I_{t}(B) \rangle =
\frac{B^2 \tau}{\tau_{\s D}^2}
\left<
	\frac{1}{
		 (b_{1x}\!\!-\!b_{2x})^2
		+ (b_{1y}\!\!-\!b_{2y})^2
		+ \frac{\tau}{\tau_{\s D}} B^2 }
\right>.
\end{equation}
The next step is averaging Eq. (\ref{deltaIav1}) over the remaining four components of the hyperfine
fields.  It is easiest to perform this integration by
switching to $\vc{b}_1 \pm \vc{b}_2$ and introducing the polar
coordinates. The integrations over the sum and over the
polar angle are elementary. The result can be cast in the
form
\begin{equation}
\label{form}
\langle \delta I_{t}(B)\rangle= \frac{1}{\tau_{\s D}}\mathcal{F}
\left(\frac{B}{B_{\s c}}\right),
\end{equation}
where the characteristic field $B_{\s c}$ is given by
\begin{equation}
\label{Bc}
B_{\s c}=\left(\frac{2\tau_{\s D}}{\tau}\right)^{1/2}b_0.
\end{equation}
The form of the function $\mathcal{F}$ is the following
\begin{equation}
\label{E1}
\mathcal{F}(x)= 2x^2 \int\limits_0^\infty du \, \frac{u}{u^2 + x^2}\, e^{-u^2}
 = x^2 e^{x^2} \text{E}_1(x^2),
\end{equation}
where $\text{E}_1(z)$ is the exponential integral function.
From Eq. (\ref{Bc}) we see that relation $\tau_{\s D} \gg \tau$ ensures that
$B_{\s c} \gg b_0$, so that the expansion Eq. (\ref{expansion}) of $\delta I_{t}(B)$ with respect to hyperfine fields
is justified.

The magnetoresistance Eq. (\ref{form}) is plotted in
Fig. \ref{figTypical}.
We note that the shape, being a single-parameter function, $\mathcal{F}(x)$,
can be very closely approximated with $x^2/(0.8+x^2)$. This approximation, which is also plotted
in Fig. \ref{figTypical},  represents
a standard fitting function for experimentally measured
magnetoresistance. It can be seen that at $x \ll 1$ there is a small deviation of $\mathcal{F}(x)$ from the approximation. This is due
to singular behavior of  $\mathcal{F}(x)$ at small arguments. This
singularity  translates into the following behavior of $\delta I_{t}(B)$
\begin{equation}
\label{behavior}
\delta I_{t}(B) \underset{b_0 < B < B_{\s c}}{\approx}
	\frac{\tau B^2}{2 \tau_{\s D}^2\,b_0^2}
	\ln \left(\frac{2 \tau_{\s D}\,b_0^2}{\tau B^2}\right).
\end{equation}
On the physical level, the fact that the ``body'' of magnetoresistance lies in
the domain $B\gg b_0$ suggests that the origin of the effect are trapping
configurations for which ${\bm \Omega}_1$ and ${\bm \Omega}_2$
are almost parallel or antiparallel. In this regard, Eqs. (\ref{form}) and (\ref{E1})
can be viewed as analytical, rather than numerical, as in Ref. \onlinecite{Bobbert1},
treatment of the bipolaron mechanism\cite{Bobbert1}.

\subsection{Averaging in the soft-pair-dominant regime }
\label{softaveraging}
Soft pairs are responsible for the second and third terms in the brackets of Eq. (\ref{duration2}) for ${\overline t}$. The second term becomes big when the sum, ${\bm \Omega}_1+{\bm \Omega}_2$, becomes anomalously small. Still it cannot dominate over the contribution from the first term
for the following reason. When ${\bm \Omega}_1+{\bm \Omega}_2$ is small, the expression in the parenthesis of the second term behaves as $({\bm \Omega}_1+{\bm \Omega}_2)^2/|{\bm \Omega}_1|^2$.
At the same time,  for small ${\bm \Omega}_1-{\bm \Omega}_2$, the expression in the
parenthesis of the first term behaves as $({\bm \Omega}_1-{\bm \Omega}_2)^2/|{\bm \Omega}_1|^2$.
In strong fields, the second expression is smaller than the first, leading to the larger $\delta I$, while in weak fields the two expressions give the same contribution to $\delta I$.

The third term in Eq. (\ref{duration2}) captures the contribution of the slow modes
to the current. Below we will study whether the averaging of this term over hyperfine
fields can dominate over the ``bipolaron" magnetic-field response given by Eq. (\ref{behavior}).

Prior to performing averaging, we rewrite the current as $I = \frac{1}{\tau_{\s D}} -
\delta I_{s}(B)$, like we did above. In the soft-pairs-dominated regime the expression
for $\delta I_{s}(B)$ takes the form
\begin{equation}
\label{correction}
\delta I_{s}(B) = \frac{1}{\tau_{\s D}}\left(
	\frac{1}{
        1 + \frac{(|\vc{\Omega}_1|^2 - |\vc{\Omega}_2|^2)^2}
            {|\vc{\Omega}_1 + \vc{\Omega}_2|^2}\tau \tau_{\s D} } \right).
\end{equation}
For a typical configuration with $|{\bm \Omega}_1|\sim |{\bm \Omega}_2|$, the
second term in denominator can be estimates as $|{\bm \Omega}_1|^2\tau \tau_{\s D}$,
so that it is large in the slow-recombination regime. This is why the soft pairs with
\begin{equation}
\label{softcondition}
(|{\bm \Omega}_1|-|{\bm \Omega}_2|)\sim \frac{1}{\sqrt{\tau\tau_{\s D}}}
\end{equation}
give
the major contribution to the average $\delta I_{s}(B)$. The latter fact allows one to simplify
the averaging procedure. Namely, one can use the fact that
 for $\epsilon \ll 1$ the combination $\frac{\epsilon}{\epsilon^2 + x^2}$ can
be replaced by  $\pi \delta(x)$. Thus, the expression to be averaged can be rewritten
in the form
\begin{equation}
\label{dI-soft}
\delta I_{s}(B) = \frac{\pi}{\tau_{\s D} \sqrt{\tau \tau_{\s D}}}
\delta \left( \frac{|\vc{\Omega}_1|^2 - |\vc{\Omega}_2|^2}
	{|\vc{\Omega}_1 +\vc{\Omega}_2|}  \right).
\end{equation}
The form Eq. (\ref{dI-soft}) suggests that characteristic magnetic field
determined from zero of the $\delta$-function is $B \sim b_{0}$, and yields the
estimate $1/\tau^{1/2}\tau_{\s D}^{3/2}b_{0}$
for $\delta I_{s}(B)$. To compare the contribution
of soft pairs to that of typical pairs this estimate should be compared
to Eq. (\ref{behavior}) taken at $B \sim b_{0}$. Soft pairs dominate
if the condition
\begin{equation}
\label{condition-soft}
 \sqrt{\frac{\tau_{\s D}}{\tau}} \gg b_{0} \tau
\end{equation}
is met. Since $\tau_{\s D}$ is much bigger than $\tau$, this condition
is compatible with the condition, $b_{ 0}\tau \gg 1$ necessary for slow recombination.
Note in passing, that replacement of the denominator in Eq. (\ref{deltaI})
by a $\delta$-function, as we did for soft pairs, is not permissible. This
follows, {\em e.g.}, from Eq. (\ref{Bc}) which suggests that the characteristic
field $B_{\s c}$ is much bigger than $b_0$. Replacement of the denominator in
Eq. (\ref{deltaI}) by a $\delta$-function would automatically fix the characteristic
field at $B \sim b_0$.

In averaging of Eq. (\ref{dI-soft}) over hyperfine configurations, we will
assume from the outset that the characteristic hyperfine fields,
$b_{1}$ and $b_{2}$,
for the electron and hole are different, so that
\begin{multline}
\label{dI-integrals}
\langle \delta I_{s}(B) \rangle =
\frac{1}{\pi^2 b_1^3 b_2^3 \sqrt{ \tau \tau_{\s D}^{3}} }
\int d^3 \vc{b}_{\s e} \int d^3 \vc{b}_{\s h} \\
\delta \left( \frac{|\vc{b}_{\s e} + \vc{B}|^2- |\vc{b}_{\s h} + \vc{B}|^2}{
	|\vc{b}_{\s e} + \vc{b}_{\s h} + 2\vc{B}|^2} \right)
	\exp\left(- \frac{|\vc{b}_{\s e}|^2}{b_1^2}
	 -\frac{|\vc{b}_{\s h}|^2}{b_2^2} \right).
\end{multline}
Subsequent analysis will indicate that different $b_{1}$ and $b_{2}$
is a necessary condition for $\delta I_{s}$ to exhibit $B$-dependence.

\begin{figure}[t]
\includegraphics[width=77mm, clip]{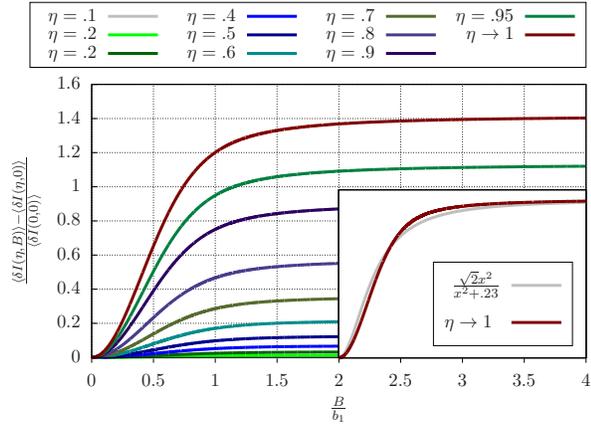}
\caption{(Color online). Magnetic field response for the ``soft-pair" mechanism is plotted from Eq. (\ref{dI-integrals5})
versus magnetic field in the units of the hyperfine field $b_{1}$  for different
values of the asymmetry parameter $\eta$. Inset: fit of the response in the limit of strong asymmetry
with conventional lineshape of OMAR,
$\sqrt{2}x^2/(0.23+x^2)$.
}
\label{figSoftI}
\end{figure}

The six-fold integral Eq. (\ref{dI-integrals})  can be reduced to
a single integral in three steps. As a first step, we introduce new variables
 $\vc{v} = \vc{b}_{\s e} - \vc{b}_{\s h}$ and
$\vc{u} = \vc{b}_{\s e} + \vc{b}_{\s h} + 2 \vc{B}$,
so that Eq. (\ref{dI-integrals}) acquires the form
\begin{multline}
\label{dI-integrals2}
\langle \delta I_{s}(B) \rangle = \frac{1}{8\pi^2 b_1^3 b_2^3 \sqrt{\tau \tau_{\s D}^2}}
\int d^3 \vc{u} \int d^3 \vc{v} \;
|\vc{u}| \, \delta(\vc{u}\cdot \vc{v}) \\
\times \exp\left(-\alpha(\vc{u}-2\vc{B})^2
+ \beta(\vc{u} - 2\vc{B})\cdot \vc{v} - \alpha |\vc{v}|^2\right),
\end{multline}
with parameters $\alpha$ and $\beta$ defined as
\begin{equation}
\label{alphabeta}
\alpha = \frac{1}{4}\left( \frac{1}{b_1^2} + \frac{1}{b_2^2} \right),~~~
\beta = \frac{1}{2} \left( \frac{1}{b_1^2} - \frac{1}{b_2^2} \right).
\end{equation}
As a second step, we perform integration over the vector ${\bm v}$.
The reason why this integration can be carried out analytically is
that, upon choosing the $z$-direction along ${\bm u}$, the $\delta$-function
fixes $v_{z}$ to be zero. The remaining two integrals over $v_{x}$ and
$v_{y}$ are simply gaussian integrals, so we get
\begin{multline}
\label{dI-integrals3}
\langle \delta I_{s}(B) \rangle = \frac{1}{8\pi b_1^3 b_2^3 \alpha
\sqrt{\tau \tau_{\s D}^3}}
\int d^3 \vc{u} \;  \\
\exp\left[ -\alpha(\vc{u} - 2 \vc{B})^2  + \frac{\beta^2}{\alpha}
\left(|\vc{B}|^2 - \frac{(\vc{B} \cdot \vc{u})^2}{|\vc{u}|^2} \right) \right].
\end{multline}
To perform the integration over ${\bm u}$, we switch to
spherical coordinates with polar axis along ${\bm B}$. Then
the integration over azimuthal angle reduces to multiplication
by $2\pi$. The third step
is the integration over the polar angle in Eq. (\ref{dI-integrals3}).
We have  
\begin{multline}
\label{dI-integrals4}
\langle \delta I_{s}(B) \rangle
	= \frac{e^{-4\alpha\left( 1 - \frac{\beta^2}{4 \alpha ^2} \right)B^2}}{
		4 b_1^3 b_2^3 \alpha \sqrt{\tau \tau_{\s D}^3}}
\int\limits_0^\infty du \; u^2 e^{-\alpha u^2} \\
\int\limits_0^\pi d\theta \sin\theta
\exp\left(4 \alpha uB \cos\theta - \frac{\beta^2}{\alpha}B^2 \cos^2\theta \right).
\end{multline}
Now we note that the integral over $\theta$ can be expressed via the error-functions in the following way
\begin{equation}
\label{erf}
\int\limits_{-1}^{1} dx\, e^{-A^2 x^2 + Cx}
 = \frac{\sqrt{\pi}}{2A} e^{\frac{C^2}{4A^2}}\!\!\!
  \left[\text{erf}\left(\!\!A\!+\!\frac{C}{2A}\!\!\right)
	 + \text{erf}\left(\!\!A\!-\!\frac{C}{2A}\!\!\right)
  \right].
\end{equation}
We are left with a single integral over $u$, which can be cast in the form
\begin{widetext}
\begin{equation}
\label{dI-integrals5}
\langle \delta I_{s}(B) \rangle
	= \frac{\sqrt{\pi}
		e^{-4\alpha\left( 1 - \frac{\beta^2}{4 \alpha ^2} \right)B^2}
	} { 8 b_1^3 b_2^3 \sqrt{\tau \tau_{\s D}^3} \sqrt{\alpha} \beta B}
\int\limits_0^\infty du \; u^2
\exp\left(- \alpha\left(1  - \frac{4\alpha^2}{\beta^2} \right)
u^2 \right)
\left[
	\text{erf}\left( \frac{\beta B}{\sqrt{\alpha}}
		+ \frac{2 \alpha^{3/2} u}{\beta} \right)
	+ \text{erf}\left( \frac{\beta B}{\sqrt{\alpha}}
		- \frac{2 \alpha^{3/2} u}{\beta} \right)
\right].
\end{equation}
\end{widetext}

\begin{figure}[t]
\includegraphics[width=77mm, clip]{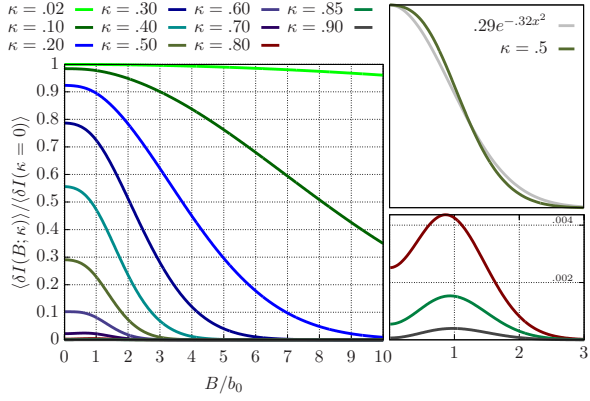}
\caption{(Color online) Magnetic field response caused by the difference,
in the $g$-factors of electron and hole is plotted from Eq. (\ref{dI-gfactors}) for
several values of relative difference, $\kappa$. Upper inset illustrates that
the shape of the response is near-gaussian. Lower inset illustrates that
at $\kappa$ close to $1$ the shape of the response develops a maximum.}
\label{figGfactors}
\end{figure}

\subsection{Analysis of Eq. (\ref{dI-integrals5})}

At this point we make an observation that for $b_{1}=b_{2}$, which is equivalent
to $\beta=0$,  magnetic
field drops out of Eq. (\ref{dI-integrals5}). The easiest way to see it
is to set $\beta=0$ at the earlier stage of calculation, namely
in Eq. (\ref{dI-integrals3})
\begin{equation}
\langle \delta I_{s}(b_{1}=b_{2}) \rangle=
\frac{1}{8 \pi b_1^3 b_2^3 \alpha \sqrt{\tau \tau_{\s D}^3}}\int d^3 \vc{u}\,
e^{-\alpha(\vc{u}-2 \vc{B})^2},
\end{equation}
which is clearly independent of $\vc{B}$ after a simple coordinate shift.
If we set $b_1 = b_2 $, then $\delta I_{s}$ is given by
\begin{equation}
\label{independent}
\langle \delta I_{s}(b_{1}=b_{2}) \rangle=
\sqrt{\frac{\pi}{2 \tau \tau_{\s D}^3}} \, \frac{1}{b_1},
\end{equation}
in agreement with the qualitative estimate above.

Magnetic field dependence of $\langle \delta I_{s} \rangle$ emerges already at
small values of asymmetry parameter defined as
\begin{equation}
\label{eta}
\eta=1-\frac{b_2^2}{b_1^2}.
\end{equation}
This is illustrated in Fig. \ref{figSoftI}, where $\langle \delta I_{s}(\eta,B)\rangle-\langle\delta I_{s}(\eta,0) \rangle$ in
the units of $\langle \delta I_{s}(\eta=0) \rangle$,  given by Eq. (\ref{independent}),
is plotted for several values of $\eta$.  We see that, as $\eta$
increases, the shape of the curves does not change much. For the saturation value the analysis
of Eq. (\ref{dI-integrals5}) yields
\begin{equation}
\label{saturation}
\frac{\langle \delta I_{s}(\eta,\infty)\rangle-\langle\delta I_{s}(\eta,0) \rangle}{
    \langle \delta I_{s}(\eta=0)\rangle}
= \frac{\sqrt{2}\eta^2}{(2-\eta)^{5/2}}.
\end{equation}
The result Eq. (\ref{dI-integrals5}) can be recast in the more concise form
in terms of the Dawson function $D(x)= e^{-x^2}\int\limits_0^x dt\, e^{t^2} $. The corresponding expression reads
\begin{equation}
\label{dawson}
\frac{\langle \delta I_s(\eta, B) \rangle}{\langle \delta I_s(0,0)\rangle}
 = \frac{\sqrt{2}}{\sqrt{2\!-\!\eta}} - \left(\!\frac{\eta}{2\!-\!\eta}\!\right)^2 \frac{\sqrt{2}}{2z}
 D\left( \frac{2z}{\sqrt{2\!-\!\eta}} \right),
\end{equation}
where we have introduced $z=B/b_1$.

In the limit of strong asymmetry, when $\eta$ is close to $1$, one gets a simple
analytical expression for  $\langle \delta I_{s}(B) \rangle$
\begin{multline}
\label{ratio}
\frac{\langle \delta I_{s}(\eta=1,B) \rangle}{\langle \delta I_{s}(\eta=0) \rangle} = \\
 2 \left(\frac{B}{b_1}\right)^2
\int\limits_{-1}^1 dx \, \sqrt{1+x} \, \exp\left[-2\left(\frac{B}{b_1}\right)^2(1-x)\right].
\end{multline}

At small $B$ the ratio Eq. (\ref{ratio}) behaves quadratically, while at large $B$
 it saturates as $\sqrt{2}\left(1 -\frac{b_1^2}{8B^2}\right)$. Overall, similarly to $I_{t}(B)$, magnetoresistance
Eq. (\ref{ratio}) can be closely approximated with $\sqrt{2}x^2/(0.23+x^2)$, as illustrated in Fig. \ref{figSoftI}.

\subsection{Inequivalence of electron and hole $g$-factors}

In the previous subsection we demonstrated that external magnetic
field drops out from the general expression Eq. (\ref{dI-integrals})
when the variances $b_1$ and $b_2$ are equal. Here we note that
averaging does not eliminate the
$B$-dependence even when $b_1=b_2$, as long as the $g$-factors of the pair partners
are different. Incorporating $g_1$ and $g_2$  into Eq. (\ref{dI-integrals})  is straightforward
 and amounts to multiplying  ${\bm b}_{\s e}+{\bm B}$ by $1+\kappa$, while
${\bm b}_{\s h}+{\bm B}$ is multiplied by $1-\kappa$, where $\kappa$ is the relative
difference in the $g$-factors. The three steps leading from Eq. (\ref{dI-integrals})
to Eq. (\ref{dI-integrals5})
are exactly the same as for $\kappa=0$. Finite $\kappa$ modifies both the prefactor
 in the integral Eq. (\ref{dI-integrals5}) and the arguments of the error functions in
  the integrand. It is convenient to analyze the magnetic
 field response by considering the ratio $\langle \delta I_{s}(B; \kappa) \rangle/
\langle \delta I_{s}(\kappa = 0) \rangle$, where the denominator is given
by Eq. (\ref{independent}).
\begin{widetext}
\begin{equation}
\label{dI-gfactors}
\frac{\langle \delta I_{s}(B; \kappa) \rangle}
{\langle \delta I_{s}(\kappa = 0) \rangle}
	= \frac{\exp\left( -\frac{2z^2}{1+\kappa^2}\right)}{2\sqrt{2}(1+\kappa^2)\kappa z}
	 \int\limits_0^\infty du \, u^2 e^{-\zeta u^2}\left[
	 	\text{erf}\left(\frac{\kappa}{1-\kappa^2}(z + \gamma u) \right)
		+ \text{erf}\left(\frac{\kappa}{1-\kappa^2}(z - \gamma u) \right)
	 \right],
\end{equation}
\end{widetext}
where $z = B/b_1$ is the scaled magnetic field.
For notational convenience we introduced the
$\kappa$-dependent
terms $\zeta$ and $\gamma$, which are defined as
\begin{equation}
\zeta = \frac{1}{2(1-\kappa^2)}\left(
1 + \frac{(1-\kappa^2)^3}{2\kappa^2(1+\kappa^2)^2}
\right),
\end{equation}
\begin{equation}
\gamma = \frac{1}{1-\kappa^2}\left( 1+ \frac{(1-\kappa^2)^3}{\kappa^2
(1+\kappa^2)}  \right).
\end{equation}
It is seen that the arguments of the error-functions
as well as the power in the exponent diverge
 in the limit $\kappa \rightarrow 1$, i.e. when the $g$-factor of one pair-partner
is zero. This divergence signifies that magnetic field
response is weak for small $(1-\kappa)$.
The underlying reason for
this is that the portion of soft pairs goes to zero if the levels of one of the
partners are not split by a magnetic field. In Fig. \ref{figGfactors} we plot the magnetic
field response for different values of $\kappa$. There are two noteworthy features
of this response. Firstly, the {\em sign} of response is opposite to that for inequivalent distributions of
electrons and holes, see Fig. \ref{figSoftI}. Secondly, the {\em shape} of $\delta I_s(B)$
is not Lorentzian anymore. In fact, this shape is close to Gaussian, as illustrated in
the inset. Another peculiar feature of $\delta I_s(B)$ which can be seen from
 Fig. \ref{figGfactors}
is that, for $\kappa$ close to $1$, the response $\delta I(B)$ develops a
bump.

\subsection{Averaging in the fast-recombination regime}
Turning to Eq. (\ref{duration3}) for $\overline{t}$ in the
fast-recombination regime we notice that the second term in the
square brackets has exactly the same form as the contribution of
the soft pairs to $\overline{t}$ in the slow-recombination regime,
see Eq. (\ref{duration2}). The underlying reason  is that, similarly
to soft pairs, this second term also comes from the slow eigenmode.
The origin of this slow eigenmode, i.e. orthogonalization of $S$-mode to
 all the other states, was discussed in detail in Sect. IIc.
Since the configurational averaging  for soft pairs was already carried out,
we conclude that the magnetic field response in the  fast-recombination regime
is simply described by Eq. (\ref{dI-integrals5}).

At this point we note that configurational
averaging over slow pairs was based on the applicability of the condition $b_{0}^2\tau\tau_{\s D}\gg 1$. Therefore, it is  important  that this condition is compatible with fast-recombination,
$b_{0}\tau \ll 1$, by virtue of a small parameter $\tau/\tau_{\s D}$.

In addition to the soft-pair contribution, Eq.  (\ref{duration3}) also contains
a term with $|{\bm \Omega}_1 \times {\bm \Omega}_2|^2$ in the denominator.
This term becomes large when ${\bm \Omega}_1$ and ${\bm \Omega}_2$ are collinear.
However, the statistical weight of these configurations is smaller than the
statistical weight of the soft-pair contribution. Indeed, in order for the term
with $|{\bm \Omega}_1 \times {\bm \Omega}_2|^2$ in denominator to become large, the angle between
the vectors ${\bm \Omega}_1$ and  ${\bm \Omega}_2$  should be restricted to
$\theta_{0} \sim 1/b_{0}\sqrt{\tau\tau_{\s D}}\ll 1$. In course of configurational
averaging, the integral, $\int d\theta \sin\theta \ldots$, emerges which is small as $\theta_{0}^2$.

We now turn to the limit of very weak hyperfine fields
for which the parameter $b_{0}^2\tau\tau_{\s D}$ is small.
One may expect that magnetic field response is suppressed in
this domain. What we demonstrate below is that this suppression
is anomalously strong. Namely, the first term of the expansion of Eq. (\ref{duration3})
with respect to  $b_{0}^2\tau\tau_{\s D}$ does not contain the external
field {\em at all}. This first term has the form

\begin{equation}
\label{fast-time}
\overline{t}-2\tau_{\s D} =
 - \frac{\tau \tau_{\s D}^2}{16}\left[
	\frac{(|\vc{\Omega}_1|^2 - |\vc{\Omega}_2|^2)^2}
		{|\vc{\Omega}_1 + \vc{\Omega}_2|^2}
	+ \frac{4 |\vc{\Omega}_1 \times \vc{\Omega}_2|^2}
		{|\vc{\Omega}_1 + \vc{\Omega}_2|^2}
\right].
\end{equation}
To realize that $B$ drops out of the expression in the square brackets
it is convenient to first replace $ |\vc{\Omega}_1 \times \vc{\Omega}_2|^2$ by
$|\vc{\Omega}_1|^2 |\vc{\Omega}_2|^2
	- (\vc{\Omega}_1 \cdot \vc{\Omega}_2)^2$ and then use the identity
\begin{equation}
\label{vectoridentity}
|\vc{\Omega}_1\!+\!\vc{\Omega}_2|^2 |\vc{\Omega}_1\!-\!\vc{\Omega}_2|^2 =
(|\vc{\Omega}_1|^2 + |\vc{\Omega}_2|^2)^2
	- 4(\vc{\Omega}_1 \cdot \vc{\Omega}_2)^2.
\end{equation}
This leads to a drastic simplification of Eq. (\ref{fast-time}), which assumes the form
\begin{equation}
\overline{t}- 2 \tau_{\s D}=
 - \frac{\tau \tau_{\s D}^2}{16} | \vc{\Omega}_1 -\vc{\Omega}_2|^2.
\end{equation}
Since $ | \vc{\Omega}_1 -\vc{\Omega}_2| = |\vc{b}_e - \vc{b}_h|$,
the magnetic field drops out of ${\overline t}$ in the first  order in $\tau\tau_{\s D}b_{0}^2$.

\section{Concluding remarks}

\begin{figure}[t]
\includegraphics[width=77mm, clip]{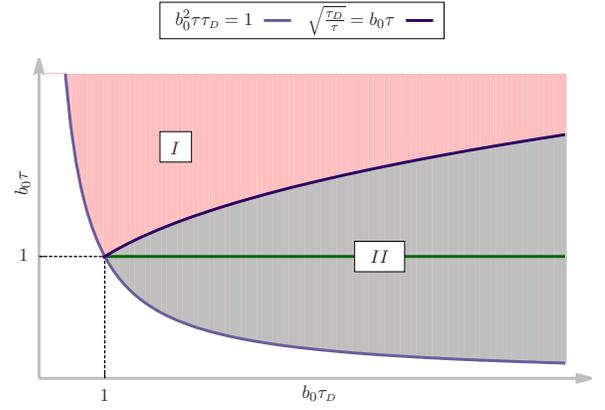}
\caption{(Color online). Different domains on the plane $\left(b_0\tau_{\s D},~ b_0\tau\right)$
illustrate the regions where different OMAR mechanisms dominate. The is no OMAR in the
white domains. The pink domain corresponds to slow recombination, and OMAR is given by Eq.
(\ref{form}). In both the upper and the lower parts of the gray domain  the OMAR
is dominated by soft pairs and is described by Eq. (\ref{dI-integrals5}). The green line divides
the gray domain into subregions where the recombination is slow (upper part) and fast (lower part).
The boundaries of the domains are: $b_0 \tau = \frac{1}{b_0 \tau_{\s D}}$, and $b_0 \tau = (b_0 \tau_{\s D})^{1/3}$.
}
\label{phase}
\end{figure}

\begin{enumerate}[(i)]
\item
Our findings can be summarized in the form of domains on the
plane $\left(b_0\tau_{\s D}, b_0\tau\right)$, as shown in Fig. \ref{phase}.
The fact that for small $b_{0}^2\tau\tau_{\s D}$ the OMAR response is absent
is reflected in Fig. \ref{phase} by leaving the domain lying below the hyperbola
uncolored. Large hyperfine fields, $b_0\tau >1$, correspond to slow recombination.
As we have demonstrated above, the OMAR for $b_0\tau >1$ can be dominated either
by ``typical" pairs or by ``soft" pairs. The corresponding regions, $I$ and $II$, are colored in Fig. \ref{phase}
by pink and gray, respectively. The domains are separated by the curve $b_0 \tau = (b_0 \tau_{\s D})^{1/3}$.
Eq. (\ref{form}) describes OMAR in the domain $I$, while in the domain $II$ Eq. (\ref{dI-integrals5}) applies.
Note that in the domain $II$ only the part above the green line corresponds to slow recombination.
The part below the green line corresponds to fast recombination, but Eq. (\ref{dI-integrals5}) applies
in both domains. The diagram describes the regimes of OMAR in low applied fields, $B\sim b_0$.
As $B$ increases above $b_0$, the gray domain shrinks.

\item
The OMAR response from the soft pairs relies exclusively on the
asymmetry between electron and hole. The evidence in favor of such an asymmetry was inferred in Ref. \onlinecite{E-H} from the analysis
of  magnetic-resonance data in organic devices.
In Ref. \onlinecite{E-H}, the ratio $b_{2}/b_{1}$ was estimated to be close to $3$, which
leads to the value of the asymmetry parameter $\eta \approx 0.9$.
Note, that bipolaron mechanism  is insensitive to the asymmetry between
electron and hole.

\item
``Parallel-antiparallel"  mechanism of Ref. \onlinecite{Bobbert1} yields the
 OMAR response  on the level of rate equations with the
transition rates calculated from the {\em golden rule}.  The applicability of this
treatment requires that the separation of Zeeman levels is large
compared to  their widths. On the other hand, the
OMAR response based on soft pairs, studied in the present paper,
comes entirely from pairs for which the Zeeman levels are almost aligned.
This  requires one to go beyond the golden rule. Previously, a similar situation
was encountered\cite{subradiance2}   by M. Schultz and F. von Oppen in
the study of transport through a nanostructure with almost degenerate levels.
The role of spin-selective recombination was played by coupling to the leads
which was strongly different for symmetric and antisymmetric combinations of the
wave functions.
M. Schultz and F. von Oppen pointed out that when two levels are closer in energy
than the width of each of them, then the conventional rate-equation-based description
is insufficient.

On the physical level, the near-degeneracy implies that some spin configuration
is preserved during many precession periods, i.e. the dynamics is important.
To account for dynamics, it is intuitively appealing to take
the result of Schulten and Wolynes, Eqs. (\ref{schultenbig})-(\ref{schultensmall}) , and multiply it by a factor describing exponential decay of population
of states due to recombination. Such an approach was adopted in
Ref. \onlinecite{Flatte1}. What this approach misses is the {\em feedback} of
recombination on the pair dynamics. It is the central message of the present paper
that this effect is strong in certain regimes, since feedback creates long-living modes.

\item

The ``parallel-antiparallel" mechanism
of Ref. \onlinecite{Bobbert1}
is based on the picture of
incoherent hopping of one of the charge carriers on the site
already occupied by the other carrier.
We considered the transport model applicable for
bipolar system where the passage of current is due
to recombination of electrons and holes.
However, the principal ingredients of both models
are the same: (a) in both transport models the spins of
the carriers precess in their effective magnetic fields, the precession
being governed by the same Hamiltonian Eq. (\ref{Hamiltonian});
(b) the passage of current
is the sequence of cycles, only one step of each cycle
is sensitive to the spin precession; (c) whether it is a hop
or recombination, it occurs only from the $S$-spin configuration;
(d) if either the hop or recombination act takes too long, the carriers
bypass each other.

\item

Both the ``parallel-antiparallel" pairs and soft pairs create
the OMAR response by blocking the current. The origin of this blocking
is completely different for the two mechanisms. In the former, the
current is blocked due to collinearity of full fields for the pair-partners,
while for the latter  the blocking is due to coincidence of their
absolute values. In general, both contributions are present in the fast-recombination
regime. The contribution of soft pairs in this regime dominates by virtue
of their statistical weight.

\item

Another distinctive feature of the soft-pairs mechanism follows from
Eq. (\ref{correction}). It contains a combination $(|\vc{\Omega}_1|^2 - |\vc{\Omega}_2|^2)^2$ in the denominator.
As the precession frequencies change with external field, ${\bm B}$,
the pair undergoes evolution from typical to soft (when $|{\bm \Omega}_1|
=|{\bm \Omega}_2|$) and back to typical. Importantly, this evolution takes place within a narrow interval of ${\bm B}$, so that at a {\em given} ${\bm B}$
only certain sparse pairs contribute to the current.
As demonstrated in Ref. \onlinecite {mesoscopics}, this redistribution
of soft pairs gives rise to {\em mesoscopic} features in $I(B)$ in small
samples.

\item
We have demonstrated above that regardless of whether the OMAR
is due to blocking caused by ``parallel-antiparallel" configurations, as in Ref. \onlinecite{Bobbert1}, or due to soft pairs, the shape of the
response is always close to $B^2/(B^2+B_{\s c}^2)$. This result was obtained
under the assumption that $\tau$ and $\tau_{\s D}$ are {\em fixed}.
If the values of $\tau$ and $\tau_{\s D}$ are broadly distributed,
then  the adequate description of transport should be based on the percolative approach\cite{Flatte1}. However, within our  minimal
model, the current is the sum of partial currents through the chains,
see Fig. \ref{figModel}. Then, with wide spread in cycle durations,  $\overline{t}$, the current  will be limited by pairs with longest $\overline{t}$ present in each chain.

\end{enumerate}

\begin{acknowledgements}
We are grateful to Z. V. Vardeny and E. Ehrenfreund for illuminating discussions.
This work was supported by NSF through MRSEC DMR-1121252 and DMR-1104495.
\end{acknowledgements}

\begin{appendix}
\section{Time Evolution and the Schrodinger Equation}

In this Appendix we sketch a formal derivation of Eqs. (\ref{generalized}) and (\ref{General})
starting from the  Liouville equation for the density operator, $\widehat{\sigma}$,
\begin{equation}
\label{liouville}
\frac{\partial \widehat{\sigma}}{\partial t} = -i [\widehat{\mathcal{H}}, \widehat{\sigma}] + \widehat{L}(\widehat{\sigma}),
\end{equation}
where the term $\widehat{L}(\widehat{\sigma})$ describes  relaxation, which in our
case is recombination from $S$ to the ground state, $G$. The ground state with energy $- \mathcal{E}$
is included into the bare Hamiltonian
\begin{align}
\widehat{\mathcal {H}} &= \widehat{H} + \widehat{H}_{G} \\
&= \left( \widehat{\vc{S}}_1 \cdot \vc{B}_1
+\widehat{\vc{S}}_2 \cdot \vc{B}_2 \right) - \mathcal{E} \ket{G}\bra{G}.
\end{align}
Then the  operator $\widehat{L}(\widehat{\sigma})$ cast into conventional Lindblad   form\cite{Lindblad} reads
\begin{equation}
\label{Lindblad}
\widehat{L}(\widehat{\sigma}) = \frac{1}{2}  \Gamma \left(
2 \ket{G}\bra{S}\widehat{\sigma}\ket{S}\bra{G}
- \widehat{\sigma}\ket{S}\bra{S} - \ket{S}\bra{S} \widehat{\sigma}
\right),
\end{equation}
where $\frac{1}{2}\Gamma=\tau^{-1}$ is the inverse recombination time.

Denote with $i$, $k$  different spin configurations of the pair prior to recombination.
The form Eq. (\ref{Lindblad}) of the dissipation ensures independence of the elements
of the density matrix with subindices $i$, $k$ from the elements containing subindex ${\s G}$.
This decoupling follows from the full system of the  equations of motion
\begin{align}
\frac{\partial \sigma_{\s GG}}{\partial t} &= \Gamma \sigma_{\s SS}, \\
\frac{\partial \sigma_{\s Gk}}{\partial t} &=
- \mathcal{E} \sigma_{\s Gk} - \sum_{i} \sigma_{\s Gi}\mathcal{H}_{\s ik}
- \frac{1}{2} \Gamma \sigma_{\s Gk} \delta_{\s Sk},
\\
\label{reduced}
\frac{\partial \sigma_{\s ik}}{\partial t} &= -i [\widehat{\mathcal{H}}, \widehat{\sigma}]_{\s ik}
 - \frac{1}{2}\Gamma \left\{ \widehat{\sigma}, \ket{S}\bra{S}  \right\}_{\s ik}.
\end{align}
Eq. (\ref{reduced}) couples only the elements of  $4\times 4$ matrix, which we denote
with $\rho$, so that Eq. (\ref{reduced}) represents equation of motion for
$\rho$. These equations can be rewritten in the form similar to
Eq. (\ref{liouville})
\begin{equation}
\label{liouville2}
\frac{\partial \widehat{\rho}}{\partial t} = -i [\widehat{H}, \widehat{\rho}] + \widehat{L}(\widehat{\rho}),
\end{equation}
with dissipation term redefined as $\widehat{L}(\widehat{\rho}) = -\frac{1}{2} \Gamma \left\{ \widehat{\rho},
\ket{S} \bra{S} \right\}$.
To derive Eq. (\ref{generalized}), we search for solution of Eq. (\ref {liouville2}) in the form
\begin{equation}
\label{trick}
\rho(t) = \ket{\psi(t)}\bra{\psi(t)},
\end{equation}
 and find that $\psi(t)$ must satisfy the following
 non-hermitian Schr\"{o}dinger equation
\begin{equation}
\label{schrodinger}
i \frac{\partial}{\partial t} \ket{\psi(t)} = \widehat{H}' \ket{\psi(t)},
\end{equation}
where $\widehat{H}'$ is defined as $\widehat{H}' =\widehat{H} - i \frac{\Gamma}{2} \ket{S}\bra{S}$.
The fact that decoupling Eq. (\ref{trick}) is valid follows from a straightforward calculation
\begin{align}
i \frac{\partial }{\partial t} \ket{\psi}\bra{\psi}
&= \left( i \frac{\partial }{\partial t} \ket{\psi} \right) \bra{\psi}
 +  \ket{\psi}\left( i \frac{\partial }{\partial t}\bra{\psi} \right) \\
 &= H' \ket{\psi}\bra{\psi} - \ket{\psi}\bra{\psi} (H')^\dagger \\
 &= \left(H - i\frac{\Gamma}{2} \ket{S}\bra{S}\right)\ket{\psi}\bra{\psi} \nonumber \\
 & \quad\quad - \ket{\psi}\bra{\psi}\left(H + i\frac{\Gamma}{2} \ket{S}\bra{S}\right) \\
 &= [H, \ket{\psi}\bra{\psi}] - i \frac{\Gamma}{2} \left\{ \ket{S}\bra{S}, \ket{\psi}
 \bra{\psi} \right\}.
\end{align}
Now Eq. (\ref{generalized}) immediately emerges as an equation for eigenvalues of the
operator $\widehat{H}'$.

\section{Derivation of Eq. (\ref{General})}
To derive Eq. (\ref{General}) for recombination time from
{\em random} initial state, we first find the expression
for recombination time, $t_{\psi_0}$, from a {\em given}
initial state, $\psi_0$, in terms of the solution of
Eq. (\ref{liouville2}) for $\rho(t)$ complemented with
condition  $\rho(0) = \ket{\psi_0}\bra{\psi_0}$.
The expression for  $t_{\psi_0}$ in terms of the
full density matrix $\widehat{\sigma}(t)$ reads
\begin{equation}
\label{dmtime}
t_{\psi_0} =  \int_0^\infty \left(dt\, \frac{\partial \sigma_{\s GG}}{\partial t}\right)t.
\end{equation}
The meaning of the expression in the brackets is the probability
that recombination took place between $t$ and $t+dt$.
The expression for  $t_{\psi_0}$ in terms of $\rho(t)$ follows
from the relation
\begin{equation}
\label{trace}
\sigma_{\s GG} + \tr\,\rho = 1.
\end{equation}
Performing integration by parts, we obtain
\begin{equation}
\label{trace1}
t_{\psi_0} =  \int_0^\infty \!\!\!\!dt\, \tr \,\rho(t).
\end{equation}
To find the
recombination time $\langle t_{\s R}\rangle$ from
the random initial state  the time $t_{\psi_0}$
should be averaged over initial states.
One way to perform this averaging
is to fix a certain
orthonormal basis, $\Phi_k$, expand $\psi_0$
as
\begin{equation}
\label{expand-basis}
\psi_0=\sum_{k}c_k\Phi_k,
\end{equation}
and  express $t_{\psi_0}$ as bilinear form in $c_k$.
This yields
\begin{align}
t_{\psi_0} &= \int_0^\infty \!\!\!\!dt \, \tr\left[
\widehat{U}(t) \psi_0^\ast \psi_0 \widehat{U}^\dagger(t) \right] \\
&=
\int_0^\infty \!\!\!\!dt \, \tr\left[
	\widehat{U}(t)\left[ \sum_k c_k \Phi_k \right]
	\left[ \sum_{k'}  {c_{k'}}^{\!\!\!\!*} {\Phi_{k'}}^{\!\!\!\!\!*} \, \right]
	\widehat{U}^\dagger(t) \right] \\
  &= \sum_{k,k'} c_k {c_{k'}}^{\!\!\!\!*} \int_0^\infty \!\!\!\!dt \,
  \tr \left[
  	\widehat{U}(t)\Phi_k {\Phi_{k'}}^{\!\!\!\!\!*}\, \widehat{U}^\dagger(t)
	\right],
\end{align}
where $\widehat{U}(t)$ is the non-unitary evolution operator.
Now the averaging over initial conditions reduces to
averaging over $c_k$ according to the rule
$\langle c_k{c_{k'}}^{\!\!\!\!*}\rangle=\frac{1}{4}\delta_{k,k'}$.
This averaging is straightforward leading to
\begin{align}
\label{sum}
\langle t_{\s R} \rangle &= \frac{1}{4} \sum_{k} \int_0^\infty \!\!\!\! dt \,
\tr \left[
	\widehat{U}(t)\Phi_k  {\Phi_{k}}^{\!\!\!\!*}\, \widehat{U}^\dagger(t)
	\right]
\end{align}
The remaining task is to express the sum,
Eq. (\ref{sum}), in terms of eigenvalues and
eigenvectors of a non-hermitian Schr\"{o}dinger equation,
Eq. (\ref{schrodinger}).
To accomplish this task we will use the expansion of the solutions $\psi_{\s k}$
of Eq. (\ref{schrodinger}), which we, for brevity, denote with $\ket{\lambda_{\s k}}$,
in terms of the orthonormal basis $\Phi_{\s k}$, which we denote with $\ket{k}$.

In terms of these new notations Eq. (\ref{schrodinger}) and the time evolution
operator can be written as
\begin{equation}
\label{time-evolve}
\widehat{H}' \ket{\lambda_j} = \lambda_j \ket{\lambda_j}, \quad
\widehat{U}(t) \ket{\lambda_j} = e^{-i \lambda_j t} \ket{\lambda_j}.
\end{equation}

It is also convenient to introduce a  matrix, $\widehat{d}$, which relates
the elements of the basis to the solutions of Eq. (\ref{schrodinger}).  Namely,
\begin{equation}
\label{d}
\ket{k} = \sum_l d_{kl} \ket{\lambda_l}.
\end{equation}
Substituting  Eq. (\ref{d}) into Eq. (\ref{time-evolve}),
we find
\begin{equation}
\label{evolution-i}
\widehat{U}(t) \ket{k}
= \sum_l d_{kl} \widehat{U}(t) \ket{\lambda_l}
= \sum_l d_{kl} e^{-i \lambda_l t} \ket{\lambda_l}.
\end{equation}
Next we introduce, $\widehat{g}$, which is the matrix
of scalar products
\begin{equation}
\label{g}
g_{ij} = \bracket{\lambda_i}{\lambda_j}.
\end{equation}
Using the definitions Eq. (\ref{d}) and Eq. (\ref{g}) we
express $\langle t_{\s R} \rangle$, defined by Eq. (\ref{sum}),
in terms of the matrices $\widehat{d}$ and $\widehat{g}$
\begin{align}
\langle t_{\s R} \rangle &= \frac{1}{4}
\sum_k \int\limits_0^\infty dt\, \tr \left[
\widehat{U}(t) \ket{k} \bra{k} \widehat{U}^\dagger(t)
\right],\\
&=
\frac{1}{4} \sum_k \int_0^\infty dt \,
\tr \left[ \widehat{U}
\left(\sum_l d_{kl} \ket{\lambda_{l}} \right)  \nonumber \right.\\
& \quad\quad \quad\quad\quad \quad\quad\quad \left.
\times \left(\sum_m d_{km}^{\ast} \bra{\lambda_{m}} \right)
\widehat{U}^\dagger
\right], \\
&=
\frac{1}{4} \sum_{k l m} \int\limits_0^\infty dt \,
e^{-i(\lambda_l - \lambda_m^\ast)t}
d_{kl} d_{km}^\ast \tr \left[ \ket{\lambda_l} \bra{\lambda_m} \right], \\
&=
\label{pre-37}
\frac{1}{4} \sum_{lm} \frac{1}{i(\lambda_l - \lambda_m^\ast)}
g_{ml} \left(\sum_k d_{kl} d_{km}^\ast \right).
\end{align}
In the last identity we have isolated the combination of the elements
of the matrix $\widehat{d}$.  The reason  is that this
combination can be cast in the form
\begin{equation}
\label{identity-d}
\sum_k d_{kl} d_{km}^\ast
= g^{-1\ast}_{ml}.
\end{equation}
To prove the latter identity, we start from the matrix relation
\begin{equation}
\bracket{\lambda_l}{i} = \sum_j d_{ij} \bracket{\lambda_l}{\lambda_j}
= \sum_j d_{ij} g_{lj},
\end{equation}
and invert it to obtain
\begin{equation}
\label{d1}
d_{ij} = \sum_l g_{jl}^{-1} \bracket{\lambda_l }{i}.
\end{equation}
Next we complex conjugate both sides of Eq. (\ref{d1}) which yields
\begin{equation}
\label{d2}
d_{ij}^\ast = \sum_l g_{jl}^{-1\ast} \bracket{\lambda_l }{i}^\ast
= \sum_l g_{jl}^{-1\ast} \bracket{i}{\lambda_l }.
\end{equation}
Now, the identity Eq. (\ref{identity-d}) emerges as a result of
straight-forward calculation
\begin{align}
\sum_k d_{kl}d_{km}^\ast &=
\sum_k \left( \sum_j g^{-1}_{lj} \bracket{\lambda_j}{k} \right)
 \left( \sum_n g^{-1\ast}_{mn} \bracket{k}{\lambda_n} \right), \\
 &= \sum_{jk}g^{-1}_{lj} g^{-1\ast}_{mn}
 \bra{\lambda_j}
 \left(\sum_k \ket{k} \bra{k} \right)
 \ket{\lambda_n} \\
&=  \sum_{jn}g^{-1}_{lj} g^{-1\ast}_{mn}g_{jn} \\
&=  g^{-1\ast}_{ml}.
\end{align}
Finally, substituting Eq. (\ref{identity-d})
into Eq. (\ref{pre-37}), we arrive at Eq. (\ref{General}) of the main
text.
\end{appendix}


\begin{thebibliography}{20}

\bibitem{frankevich} E. L. Frankevich, I. A. Sokolik, D. I. Kadyrov, and V. M. Kobryanskii, Pis'ma
Zh. Eksp. Teor. Fiz. {\bf 36}, 401 (1982) [JETP Lett. {\bf 36}, 488 (1982)].

\bibitem{frankevich1} E. L. Frankevich, A. A. Lymarev, I. Sokolik, F. E. Karasz, S. Blumstengel, R. H. Baughman,
and H. H. H{\"o}rhold, Phys. Rev. B {\bf 46}, 9320 (1992).

%{Markus0,Markus1,Markus2,Prigodin,Valy0,Valy1,Valy2,Valy3,Bobbert1,Bobbert2,Wagemans2,Wagemans,Wagemans1,Flatte1}
\bibitem{reviews}
%See, e.g.,
K. M. Salikhov, Y. N. Molin, R. Z. Sagdeev, and A. L. Buchachenko, in {\em Spin Polarization and
Magnetic Effects in Radical Reactions}, edited by Y. N. Molin (Elsevier, Amsterdam, 1984), pp. 32-116, and the review U. E. Steiner and T. Ulrich, Chem. Rev. {\bf 89}, 51 (1989).

\bibitem{Schulten}
K. Schulten and P. G. Wolynes, J. Chem. Phys. {\bf 68}, 3292 (1978).

\bibitem{Markus0}
T. L. Francis, {\"O}. Mermer, G. Veeraraghavan, and M. Wohlgenannt,
New J. Phys. {\bf 6}, 185 (2004).

\bibitem{Markus1}
{\"O}. Mermer, G. Veeraraghavan, T. L. Francis, Y. Sheng, D. T. Nguyen, M. Wohlgenannt, A. Köhler, M. K. Al-Suti, and M. S. Khan, Phys. Rev. B {\bf 72}, 205202 (2005).
%Large magnetoresistance in nonmagnetic ?-conjugated semiconductor thin film devices

\bibitem{Markus2} Y. Sheng, T. D. Nguyen, G. Veeraraghavan, O. Mermer, M. Wohlgenannt, S. Qiu, and U. Scherf, Phys. Rev. B {\bf 74}, 045213 (2006).
%Hyperfine interaction and magnetoresistance in organic semiconductors

\bibitem{Markus3} F. Wang, F. Maci{\'a}, M. Wohlgenannt, A. D. Kent, and M. E. Flatt{\'e}, Phys. Rev. X {\bf 2}, 021013 (2012).

\bibitem{Gillin} P. Desai, P. Shakya, T. Kreouzis, and W. P. Gillin, Phys. Rev. B {\bf 76}, 235202 (2007); Sijie Zhang, N. J. Rolfe, P. Desai, P. Shakya, A. J. Drew, T. Kreouzis, and W. P. Gillin, Phys. Rev. B {\bf 86}, 075206 (2012).

\bibitem{Valy0} F. J. Wang, H. B{\"a}ssler, and
Z. Valy Vardeny, Phys. Rev. Lett. {\bf 101}, 236805 (2008).
%Magnetic Field Effects in ?-Conjugated Polymer-Fullerene Blends:
%Evidence for Multiple Components

\bibitem{Valy1}
T. D. Nguyen, G. Hukic-Markosian, F. Wang, L. Wojcik, X.-G. Li,
E. Ehrenfreund, and Z. V. Vardeny,
Nat.  Mater. {\bf 9}, 345 (2010).

\bibitem{Valy2} T. D. Nguyen, B. R. Gautam, E. Ehrenfreund, and Z. V. Vardeny, Phys. Rev. Lett. {\bf 105}, 166804 (2010).

\bibitem{Valy3} Tho D. Nguyen, T. P. Basel, Y.-J. Pu, X-G. Li, E. Ehrenfreund,
and Z. V. Vardeny, Phys. Rev. B {\bf 85}, 245437 (2012).
%Magnetic field effect on excited-state spectroscopies of
%?-conjugated polymer films

\bibitem{Blum}
F. L. Bloom, W. Wagemans, M. Kemerink, and B. Koopmans,
Phys. Rev. Lett. {\bf 99}, 257201 (2007).
%Separating Positive and Negative Magnetoresistance in Organic Semiconductors,

\bibitem{Bobbert1} P. A. Bobbert, T. D. Nguyen, F. W. A. van Oost, B. Koopmans, and M. Wohlgenannt,  Phys. Rev. Lett. {\bf 99}, 216801 (2007).
%Bipolaron Mechanism for Organic Magnetoresistance,

\bibitem{Bobbert2} W. Wagemans, F. L. Bloom, P. A. Bobbert, M. Wohlgenannt, and B. Koopmans, J. Appl. Phys. {\bf 103}, 07F303 (2008).
%A Two-Site Bipolaron Model for Organic Magnetoresistance,

\bibitem{Wagemans2}
F. L. Bloom, M. Kemerink, W. Wagemans, and B. Koopmans, Phys. Rev. Lett. {\bf 103}, 066601 (2009).
%Sign Inversion of Magnetoresistance in Space-Charge Limited Organic Devices

\bibitem{Wagemans}
S. P. Kersten, A. J. Schellekens, B. Koopmans, and P. A. Bobbert, Phys. Rev. Lett. {\bf 106}, 197402 (2011).
%Magnetic-Field Dependence of the Electroluminescence of Organic Light-Emitting Diodes: A %Competition between Exciton Formation and Spin Mixing

\bibitem{Wagemans1} W. Wagemans, A. J. Schellekens, M. Kemper,
F. L. Bloom, P. A. Bobbert, and B. Koopmans
Phys. Rev. Lett. {\bf 106}, 196802 (2011).
%Spin-Spin Interactions in Organic Magnetoresistance Probed by Angle-Dependent Measurements

\bibitem{WagReview} W. Wagemans and B. Koopmans, Phys. Status Solidi B {\bf 248}, 1029
    (2011).

\bibitem{Prigodin} V. N. Prigodin, J. D. Bergeson, D. M. Lincoln,
and A. J. Epstein, Synth. Met. {\bf 156}, 757 (2006).
%Anomalous Room Temperature Magnetoresistance in Organic Semiconductor, Synth. Met. 156, 757 (2006).

\bibitem{Flatte1} N. J. Harmon and M. E. Flatt{\'e}, Phys. Rev. Lett. {\bf 108}, 186602 (2012);
Phys. Rev. B {\bf 85}, 075204 (2012); Rev. B {\bf 85}, 245213 (2012).
%Spin-Flip Induced Magnetoresistance in Positionally Disordered Organic Solids

\bibitem{Dicke}  R. H. Dicke, Phys. Rev. {\bf 93}, 99 (1954).

\bibitem{subradiance1} T. V. Shahbazyan and M. E. Raikh, Phys. Rev. B {\bf 49}, 17123 (1994).

\bibitem{subradiance2}  M. G. Schultz and F. von Oppen, Phys. Rev. B {\bf 80}, 033302 (2009).

\bibitem{Gefen} J. K\"{o}nig, Y. Gefen, and G. Sch\"{o}n,
Phys. Rev. Lett. {\bf 81}, 4468 (1998).

\bibitem{E-H} D. R. McCamey, K. J. van Schooten, W. J. Baker, S.-Y. Lee, S.-Y. Paik, J. M. Lupton, and C. Boehme,
Phys. Rev. Lett. {\bf 104}, 017601 (2010);
S.-Y. Lee, S.-Y. Paik, D. R. McCamey, J. Yu, P. L. Burn, J. M. Lupton, and C. Boehme, J. Am. Chem. Soc. {\bf 133}, 072019 (2011).

\bibitem{mesoscopics} R. C. Roundy, Z. V. Vardeny, M. E. Raikh,
e-print arXiv:1210.3443v1 (2012).

\bibitem{Lindblad}  G. Lindblad, Commun. Math. Phys. {\bf 48}, 119 (1976).
\end{thebibliography}
\end{document}